\documentclass[english,prl,twocolumn,notitlepage,superscriptaddress] {revtex4-1}
\pdfoutput=1
\usepackage{graphicx}
\usepackage{amssymb}
\usepackage{setspace}
\usepackage{epstopdf}
\usepackage{multirow}
\usepackage{bm}
\usepackage{braket}
\usepackage{amsmath}

\renewcommand{\vec}[1]{\bm{#1}}

\usepackage{color}

\newcommand\ymn[1]{\textcolor{black}{#1}}

\begin{document}

\title{Electrical spin driving by $g$-matrix modulation in spin-orbit qubits}

\author{Alessandro Crippa}
\email{alessandro.crippa@cea.fr}
\affiliation{Universit\'e Grenoble Alpes \& CEA INAC-PHELIQS, F-38000 Grenoble, France}

\author{Romain Maurand}
\affiliation{Universit\'e Grenoble Alpes \& CEA INAC-PHELIQS, F-38000 Grenoble, France}

\author{L\'eo Bourdet}
\affiliation{Universit\'e Grenoble Alpes \& CEA INAC-MEM, F-38000 Grenoble, France} 

\author{Dharmraj Kotekar-Patil}
\affiliation{Universit\'e Grenoble Alpes \& CEA INAC-PHELIQS, F-38000 Grenoble, France}

\author{Anthony Amisse}
\affiliation{Universit\'e Grenoble Alpes \& CEA INAC-PHELIQS, F-38000 Grenoble, France}

\author{Xavier Jehl}
\affiliation{Universit\'e Grenoble Alpes \& CEA INAC-PHELIQS, F-38000 Grenoble, France}

\author{Marc Sanquer}
\affiliation{Universit\'e Grenoble Alpes \& CEA INAC-PHELIQS, F-38000 Grenoble, France}

\author{Romain Lavi\'eville}
\affiliation{Universit\'e Grenoble Alpes \& CEA LETI MINATEC campus, F-38000 Grenoble, France}

\author{Heorhii Bohuslavskyi}
\affiliation{Universit\'e Grenoble Alpes \& CEA INAC-PHELIQS, F-38000 Grenoble, France}
\affiliation{Universit\'e Grenoble Alpes \& CEA LETI MINATEC campus, F-38000 Grenoble, France}

\author{Louis Hutin}
\affiliation{Universit\'e Grenoble Alpes \& CEA LETI MINATEC campus, F-38000 Grenoble, France}

\author{Sylvain Barraud}
\affiliation{Universit\'e Grenoble Alpes \& CEA LETI MINATEC campus, F-38000 Grenoble, France}

\author{Maud Vinet}
\affiliation{Universit\'e Grenoble Alpes \& CEA LETI MINATEC campus, F-38000 Grenoble, France}

\author{Yann-Michel Niquet}
\email{yniquet@cea.fr}
\affiliation{Universit\'e Grenoble Alpes \& CEA INAC-MEM, F-38000 Grenoble, France} 

\author{Silvano De Franceschi}
\email{silvano.defranceschi@cea.fr}
\affiliation{Universit\'e Grenoble Alpes \& CEA INAC-PHELIQS, F-38000 Grenoble, France}


\begin{abstract}
In a semiconductor spin qubit with sizable spin-orbit coupling, coherent spin rotations can be driven by a resonant gate-voltage modulation. Recently, we have exploited this opportunity in the experimental demonstration of a hole spin qubit in a silicon device. Here we investigate the underlying physical mechanisms by measuring the full angular dependence of the Rabi frequency as well as the gate-voltage dependence and anisotropy of the hole $g$ factors. We show that a $g$-matrix formalism can simultaneously capture and discriminate the contributions of two mechanisms so far independently discussed in the literature: one associated with the modulation of the $g$ factors, and measurable by Zeeman energy spectroscopy, the other not. Our approach has a general validity and can be applied to the analysis of other types of spin-orbit qubits.
\end{abstract}

\maketitle

The spin-orbit (SO) interaction is a relativistic effect coupling the motional and spin degrees of freedom of a particle. In semiconductors, a sufficiently strong SO coupling can allow for electric-dipole spin resonance (EDSR) \cite{RashbaBook, rashbaReview}, defined as the coherent rotation on an electron spin driven by a radio-frequency electric field. This opportunity can be exploited for the control of semiconductor spin qubits, where quantum information is encoded in the spin state of a confined electron (or hole). As opposed to magnetically driven spin resonance \cite{Koppens, pla_electronqubit, VeldhorstAddressable}, EDSR can result in faster manipulation \cite{YonedaPRL2014,roro} and higher qubit fidelity \cite{Yoneda2017}.\\
Two distinct mechanisms for EDSR ultimately related to SO coupling have been identified so far. The first one, known as $g$-tensor magnetic resonance ($g$-TMR), was originally observed in a GaAs/AlGaAs heterostructure \cite{Kato}. This mechanism requires anisotropic and spatially varying electron \cite{Kato, FlattePRB,PettaGtuning, TaruchaPRBR13} (or hole \cite{NataliaAPL,VoisinNL}) $g$ factors. \ymn{In essence, an alternating electric field modulates the confinement potential and the shape of the electron wave function, which translates into time-dependent $g$ factors and, therefore, into a non-collinear modulation of the Larmor vector.} The second mechanism was experimentally observed in a variety of III-V semiconductor quantum dots \cite{Nowack07, LPKSOqubit, PeterssonQED, SOfastLPK}. It is not associated with a modulation of the $g$ factors. \ymn{Instead, the alternating electric field shakes the wave function as a whole. During this motion, SO interactions give rise to} an effective time-dependent magnetic field proportional to the alternating electric field, to the static magnetic field, and to the inverse spin-orbit length \cite{Levitov, Debald, FlensbergSO, Golovach}.\\
In general, the two mechanisms are expected to coexist, since they share a common SO origin. Here we investigate this coexistence in a hole spin qubit confined to a silicon quantum dot. We induce Rabi oscillations of a spin-1/2 hole state by means of a gate-voltage RF modulation. By correlating the anisotropy of the Rabi frequency with the angular and gate-voltage dependence of the $g$ factors, we discriminate the mechanisms contributing to EDSR.\\
\ymn{To this aim, we relate EDSR to the gate-voltage modulation of a $g$-matrix $\hat{g}$. We show that part of this
modulation can be reconstructed from the gate-voltage and magnetic field dependence of the $g$ factor, which is defined as $|g^*|=|\hat{g}\cdot\vec{b}|$, where $\vec{b}=\vec{B}/|\vec{B}|$ is the unit vector pointing along the magnetic field $\vec{B}$. This contribution can be cast as a generalization of the $g$-TMR mechanism discussed in Ref.\,\cite{Kato}. Unitary modulations of the $g$-matrix, which do not give rise to variations of the $g$ factor, can induce Rabi oscillations too. The EDSR mechanism discussed in Ref.\,\cite{Golovach} is an example of this scenario. To emphasize the absence of Zeeman energy modulations, we shall refer to such contributions as iso-Zeeman EDSR (IZ-EDSR). Remarkably, they can only be extracted from time-domain measurements of the Rabi frequency as a function of the orientation of $\vec{B}$. 
\\Finally, our work shows that the electrical driving of a spin-orbit qubit can be fully characterized by measuring the $g$ factors and Rabi frequencies for a few magnetic field orientations. Besides bringing a deeper understanding of the underlying physics, our results provide useful guidelines to improve device design and find operating conditions for optimal qubit driving.}
\begin{figure}
\centering
	\includegraphics[width=\columnwidth]{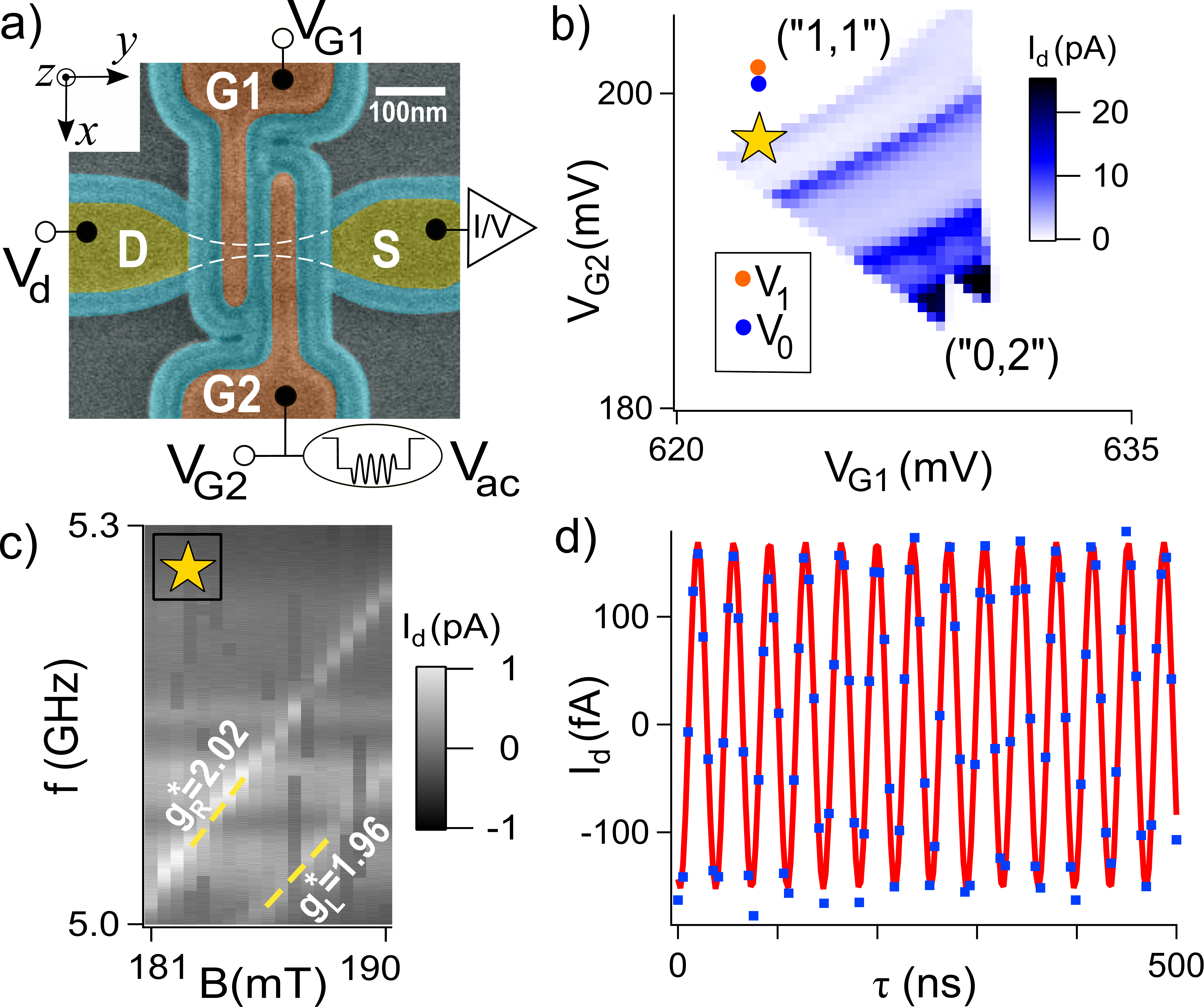} 
	\caption{(Color online) a) Scanning electron micrograph of a device similar to the one measured, with false colors. The white dashed lines outline the Si nanowire, 25\,nm large and 8\,nm thick, extending between S and D (yellow). Gates G1, G2 (brown) are 35\,nm long and separated by Si$_3$N$_4$ spacers (cyan); microwave pulses are applied to G2. b) Bias triangles of the double dot under study at $V_{d}=5$ mV. The quoted numbers denote the equivalent excess charges inferred from Pauli blockade; $V_0$, $V_1$ mark points where spins are manipulated during pulse sequences. c) Spectroscopy of the spin-orbit qubit under Pauli spin blockade [yellow star of panel b)]; $I_d$ is plotted as a function of microwave frequency $f$ and $|\vec{B}=(B,0,0)|$. The slope of the resonance lines gives $g^*_L$ and $g^*_R$. d) Rabi oscillations as a function of microwave burst time $\tau$ for $\vec{B}=(0,0,-0.36)$\,T. A current offset of 200\,fA is subtracted for clarity.}
	\label{fig:Fig1}
\end{figure}
\\The device, shown in Fig.\,\ref{fig:Fig1}a, is a double gate metal-oxide-semiconductor (MOS) silicon nanowire field-effect transistor (FET) \cite{afsid}. The source (S) and drain (D) are degenerately boron-doped reservoirs of holes, whereas the channel, oriented along [110], is undoped. Two top gates in series (G1, G2) tune a double quantum dot at the base temperature $T=15$\,mK of a dilution cryostat. At finite drain bias $V_{d}$, the transport of holes through the double dot results in pairs of triangles of DC current $I_{d}$ in the gate voltage diagram $V_{G1}$-$V_{G2}$. We work in the region of Fig.\,\ref{fig:Fig1}b, where Pauli spin blockade \cite{TaruchaPSB} is revealed by current rectification at the base of the triangles and by magnetotransport measurements \cite{DharamPSB, HeorhiiPSB, HamiltonNL15}; (``1,\,1") and (``0,\,2'') denote the parity-equivalent excess charges of the double dot, though each dot contains between 10 and 30 holes, as in Ref.\,\cite{roro}.\\
In Fig.\,\ref{fig:Fig1}c we operate the device as a spin-orbit qubit. The gates are biased in Pauli spin blockade (yellow star in Fig.\,\ref{fig:Fig1}b), while a continuous microwave signal of frequency $f$ is applied to G2 with $\vec{B}=(B,0,0)$. A microwave excitation drives hole spin transitions in both quantum dots, thereby lifting Pauli blockade. $I_{d}$ increases when the photon energy $hf$ matches the Zeeman splitting $\Delta E=|g^*|\mu_B B$ between two spin states, $h$ being Planck's constant, $\mu_B$ Bohr's magneton and $g^*$ the effective hole $g$ factor of the resonant dot for this orientation of $\vec{B}$. From Fig.\,\ref{fig:Fig1}c we obtain $|g^*_L|=1.96 \pm 0.02$ and $|g^*_R|=2.02 \pm 0.02$ in the left and right dot, respectively.\\
Figure \ref{fig:Fig1}d shows Rabi oscillations of the spin in dot R. The spin is initialized by Pauli blockade at the yellow star in Fig.\,\ref{fig:Fig1}b. The coherent manipulation is accomplished by a microwave burst of duration $\tau$ applied in the Coulomb blockade regime at the bias point $V_0$ marked by a blue dot in Fig.\,\ref{fig:Fig1}b. Spin readout relies again on spin blockade back at the yellow star. The initialization-manipulation-readout sequence is repeated continuously, resulting in an oscillating current $I_d$ as a function of $\tau$, a signature of the coherent spin rotations \cite{Nowack07, roro}.
\begin{figure}
\centering
	\includegraphics[width=\columnwidth]{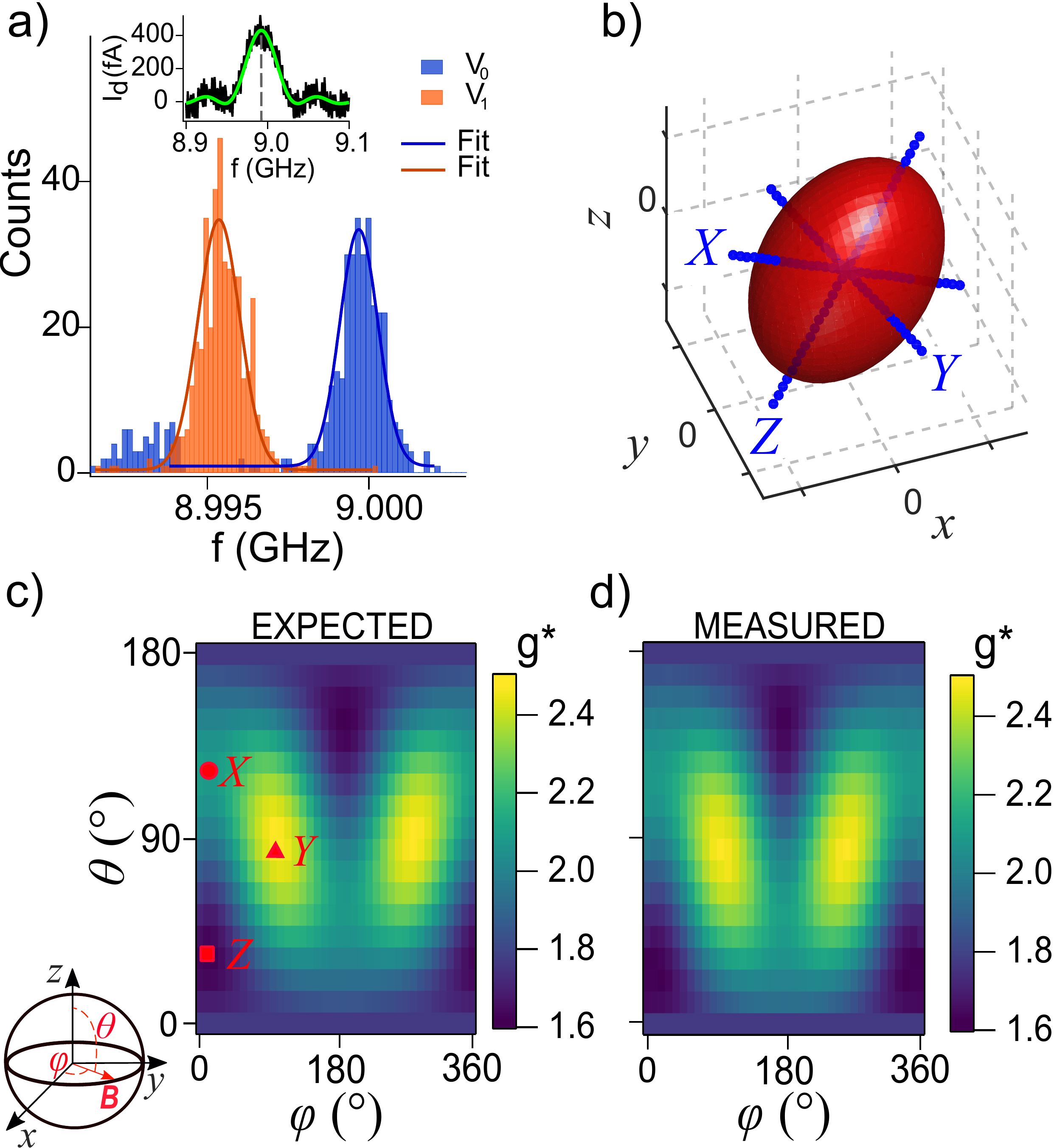} 
	\caption{(Color online) a) Inset: current trace as a function of microwave frequency $f$ for burst duration $\tau=20$\,ns at $\vec{B}=(0, 216, 216)$\,mT. We fit with $I_d \propto (f_r^2/a) \cdot \sin^2(\pi\tau \sqrt{a})$, where $a=(f-f_0)^2+f_r^2$, with $f_r, f_0=|g^*|\mu_B B/h$ the Rabi and resonance frequency, respectively \cite{Vandersypen_electronqubit}. Blue and orange histograms display the dispersion of $f_0$ when $V_{G2}$ is pulsed to point $V_0$ and $V_1$, respectively. The Gaussian fits yield $g^*(V_0)=2.013 \pm 0.001$ and $g^*(V_1)=2.010 \pm 0.001$ (including the field uncertainty). b) Typical isosurface of $\Delta E^2=g^{*2}\mu_B^2B^2$ in the measurement frame $\{\vec{x},\,\vec{y},\,\vec{z}\}$. The blue lines are the principal magnetic axes $\vec{X},\,\vec{Y},\,\vec{Z}$ of the ellipsoids at $V_{G2}=V_0$. c) $g^*$ as a function of the field angles $\theta, \varphi$, reconstructed from six values of $g^*$ uniquely defining the ellipsoids. d) Experimental cartography of $g^*$. $\theta, \varphi$ are stepped by $10^{\circ}$ between $0^{\circ}$ and $180^{\circ}$. The whole plot is obtained by symmetry ($\vec{B} \to -\vec{B}$).}
	\label{fig:Fig2}
\end{figure}
\\\ymn{To elucidate the origin of the observed EDSR, we investigate the anisotropy of the hole $g$ factor and Rabi frequency in dot R at bias point $V_0$. For that purpose, we map dot R onto an effective two-level system $\ket{\Uparrow}$ and $\ket{\Downarrow}$. To first order in $\vec{B}$, the effective spin Hamiltonian of this system reads}
\begin{equation}
H=\frac{1}{2}\mu_B{^t}\vec{\sigma}\cdot\hat{g}\cdot\vec{B}\,,
\label{eq:gtensor}
\end{equation}
\ymn{where $\vec{\sigma}=(\sigma_1,\sigma_2,\sigma_3)$ are the Pauli matrices. $H$ is fully parametrized by the nine independent elements of the matrix $\hat{g}$. For a given $\vec{B}$, the square of the Zeeman splitting between the eigenstates of $H$ is:
\begin{equation}
\Delta E^2 = |g^*|^2\mu_B^2 B^2 = \mu_B^2|\hat{g}\cdot\vec{B}|^2 = \mu_B^2({^t}\vec{B}\cdot\hat{G}\cdot\vec{B})\,,
\label{eq:deltaEsquare}
\end{equation}
where $\hat{G}={^t}\hat{g}\cdot\hat{g}$ is the symmetric Zeeman tensor. The eigenvectors of $\hat{G}$ are the so-called principal magnetic axes $\vec{X}$, $\vec{Y}$, $\vec{Z}$, while the corresponding eigenvalues $g_1^{*2}$, $g_2^{*2}$ and $g_3^{*2}$ are the squares of the principal $g$ factors. $\hat{G}$ completely characterizes the variations of $|g^*|$, and can be uniquely reconstructed from the measurement of $\Delta E$ for six orientations of $\vec{B}$ (while $\hat{g}$ can be reconstructed only up to a unitary transform leaving $|\hat{g}\cdot\vec{B}|$ invariant).} For each orientation, we measure $I_d$ as a function of $f$ for a fixed burst time while applying the sequence of gate pulses described above. The current trace shows a peak at the resonance frequency $hf=|g^*|\mu_B B$ (inset of Fig.\,\ref{fig:Fig2}a). Such a measurement is repeated 400 times in order to get the resonance frequency distribution shown in Fig.\,\ref{fig:Fig2}a. Finally $|g^*(V_0, \vec{B})|$ is extracted from the peak of the Gaussian distribution.\\
We find $|g^*_1|\simeq 2.08$, $|g^*_2|\simeq 2.48$ and $|g^*_3|\simeq 1.62$. Figure\,\ref{fig:Fig2}b shows the ellipsoidal isosurfaces of $\Delta E^2$ in the measurement frame $\{\vec{x},\,\vec{y},\,\vec{z}\}$. Figures\,\ref{fig:Fig2}c and\,\ref{fig:Fig2}d compare the full angular dependence of $g^*$ reconstructed from  the measurement of $\Delta E$ for six orientations of $\vec{B}$ and Eq.\,(\ref{eq:deltaEsquare}), to the experimental values as a function of the elevation angle $\theta$ and azimuthal angle $\varphi$ of $\vec{B}$ in the measurement frame. Each pixel of Fig.\,\ref{fig:Fig2}d is a distinct spin resonance experiment. The magnetic axis $\vec{Y}$ associated to $g^*_2$ is almost aligned with the $\vec{y}$ (nanowire) axis. The other two orthogonal magnetic axes do not match crystallographic axes.\\ 
We have measured similar in-plane $g^*\sim 2-2.6$ (depending on the gate voltage) and out-of-plane $g^*\sim 1.5$ in other $p$-type nanowire FETs \cite{VoisinNL}. The fact that the ratio between the highest and lowest $g$ factors is not very large suggests a significant heavy/light-hole mixing \cite{Winkler03, HamiltonNL15}.
\\The $g$ factors can also be tuned by the electric field. An example is the orange resonance frequency distribution of Fig.\,\ref{fig:Fig2}a, measured at another manipulation point $V_1$ (orange dot in Fig.\,\ref{fig:Fig1}b) slightly shifted from $V_0$. Similar changes for other orientations of $\vec{B}$ indicate a small (yet sizable) rotation of the principal magnetic axes.
\begin{figure}
\centering
	\includegraphics[width=\columnwidth]{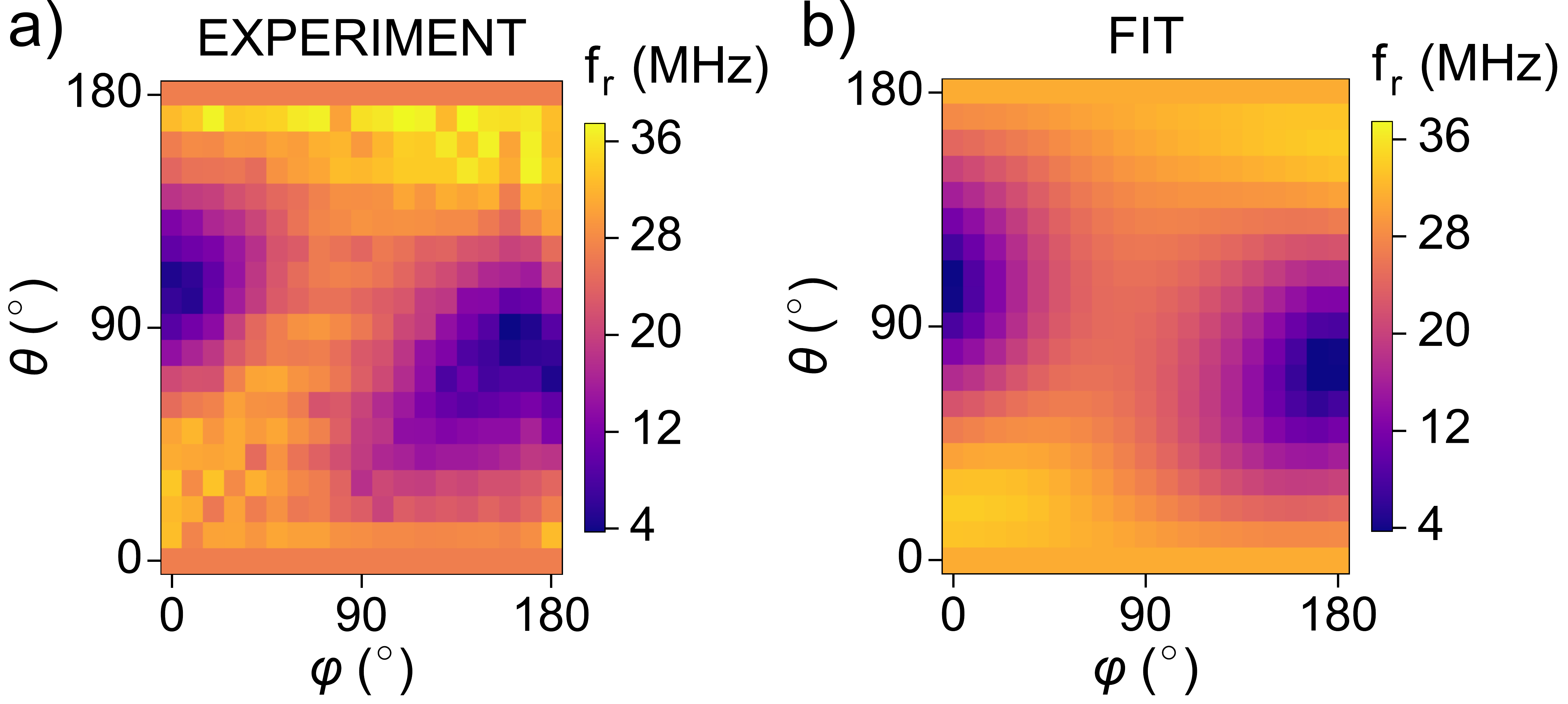} 
	\caption{(Color online) a) Angular dependence of the Rabi frequency. The two strong minima [($\varphi\approx0^{\circ}$, $\theta\approx90^{\circ}$) and ($\varphi\approx180^{\circ}$, $\theta\approx90^{\circ}$)] suggest that the spin-orbit field $\vec{B}_{\rm SO}$ is in-plane, perpendicular to nanowire axis (along $\vec{x}$). The map is recorded at constant Zeeman splitting, see main text. b) Analogous map obtained from the fit of a) with Eq.\,(\ref{eq:Rabi}).}
	\label{fig:Fig3}
\end{figure}
\\In addition to the Zeeman tensor, we characterize the dependence of the Rabi frequency $f_r$ on the orientation of $\vec{B}$. Figure\,\ref{fig:Fig3}a displays $f_r$ for 291 directions of $\vec{B}$. For each pixel, $|\vec{B}|$ is adjusted so that the spin resonance sticks to 9 GHz to drive the spin always at the same microwave power. The measured $f_r$ ranges from 3 to 40 MHz; it is maximal for $\vec{B} \parallel \vec{z}$, while it is minimal along the $\vec{x}$ direction.
%
\\\ymn{To disentangle the mechanisms for Rabi oscillations, we intend to draw a unified picture for the electrical  tunability of $g^*$ (Figs.\,\ref{fig:Fig2}a,\,\ref{fig:Fig2}d) and the anisotropy of the Rabi map (Fig.\,\ref{fig:Fig3}a). Our approach consists in a generalization of the formalism of Ref.\,\cite{Kato}. We first expand $\hat{g}(V_G)\simeq\hat {g}(V_0)+(V_G-V_0)\,\hat{g}^\prime(V_0)$ to first order in $V_G$, with $\hat{g}^\prime=\partial\hat{g}/\partial V_G$. Then, upon application of a microwave burst $V_G(t)=V_0+V_{ac}\sin(2\pi ft)$ resonant with the Kramers doublet transition ($hf=\Delta E$), the Rabi frequency $f_r$ is the norm of the Rabi vector \cite{NataliaAPL,SupplMat}
\begin{equation}
\vec{f}_r=\frac{\mu_B B V_{ac}}{2h|g^*|} \Bigl[ \hat{g}(V_0)\cdot\vec{b}\Bigr] \times \Bigl[ \hat{g}^\prime(V_0)\cdot\vec{b} \Bigr]\,.
\label{eq:Rabi}
\end{equation}
Any linear-in-$B$ and $V_{ac}$ mechanism for the Rabi oscillations can be mapped onto Eq.\,(\ref{eq:gtensor}) with a $V_G$-dependent $\hat{g}$ matrix and must be captured by Eq.\,(\ref{eq:Rabi}). Still, the $g$ factor anisotropy (characterized by $\hat{G}$) and its dependence on $V_G$ (characterized by $\hat{G}^\prime=\partial\hat{G}/\partial V_G$) do not provide enough information to reconstruct $\hat{g}$ and $\hat{g}^\prime$ and make use of Eq.\,(\ref{eq:Rabi}) to predict Rabi frequencies. As discussed above, $\hat{g}$ is defined by the experimental $\hat{G}$ only up to a unitary transform, or equivalently up to a choice for the Kramers basis $\{\ket{\Uparrow}, \ket{\Downarrow}\}$. There always exists a Kramers basis in which $\hat{g}(V_0)\equiv{\rm diag}(g_1^*,g_2^*,g_3^*)$ is diagonal in the magnetic axes frame $\{\vec{X},\vec{Y},\vec{Z}\}$ \cite{ChibotaruPRL08,SupplMat,Weil-Bolton}. However, it is usually not possible to reconstruct $\hat{g}^\prime(V_0)$ from $\hat{G}^\prime(V_0)$ in the same basis set, because the latter remains unknown.}\\
We emphasize that a gate-voltage modulation of $\hat{g}$ can give rise to a finite $f_r$ [Eq.\,(\ref{eq:Rabi})] but no variations in the Zeeman splitting, i.e. $\hat{G}^\prime=0$. Indeed, $\hat{G}^\prime$ is related to the derivative of the $\hat{g}$ matrix by
\begin{equation}
\hat{G}^\prime={}^{t}\hat{g}\cdot\hat{g}^\prime+{}^t\hat{g}^\prime\cdot\hat{g}\,,
\label{eq:deltaG}
\end{equation}
so that $\hat{G}^\prime$ is zero if ${}^{t}\hat{g}\cdot\hat{g}^\prime$ is an antisymmetric matrix. \ymn{We refer to the contributions from such modulations as ``Iso-Zeeman EDSR'' (IZ-EDSR).} A notable example is the set-up of Ref.\,\cite{Golovach}, where a harmonic quantum dot in a static magnetic field is moved around its equilibrium position by an homogeneous alternating electric field, and EDSR is mediated by intrinsic SO coupling. In these conditions, $\hat{G}^\prime=0$ because the electric field does not change the shape of the confinement potential and, therefore, the Zeeman splitting; however, $\hat{g}^\prime\ne0$ since the vector potential breaks translational symmetry. The $g$-matrix formulation of the theory of Ref.\,\cite{Golovach} for harmonic and arbitrary confinement potentials is discussed in \cite{SupplMat}.\\
On the other hand, the electrical variations $\hat{G}^\prime$ of the Zeeman tensor result in a nonzero $\hat{g}^\prime$ [Eq.\,(\ref{eq:deltaG})] driving spin rotations according to Eq.\,(\ref{eq:Rabi}). \ymn{This contribution to EDSR is a generalization of the $g$-TMR mechanism. In the original $g$-TMR scenario of Refs.\,\cite{Kato, NataliaAPL}, the spin oscillations result only from modulations of the principal $g$ factors $g_i^*$, but the magnetic axes are imposed by the symmetries of the system and do not depend on the gate voltages.}\\
Both IZ-EDSR and $g$-TMR are ultimately due to the SO interaction, yet they manifest in different ways. $g$-TMR is driven by modulations of the shape of the confinement potential by the electric field, which result in variations of the Zeeman splittings. IZ-EDSR also catches the effect of the alternating motion of the dot as a whole (as, e.g., in Ref.\,\cite{Golovach}), which is invisible in the Zeeman splitting.\\
Our formalism allows to discriminate IZ-EDSR and $g$-TMR within Fig.\,\ref{fig:Fig3}a. The electrical modulations of $\hat{g}$ due to $g$-TMR and IZ-EDSR can be characterized by two matrices, $\hat{g}_{\rm TMR}^\prime$ and $\hat{g}_{ \rm IZR}^\prime$, such that $\hat{g}_{\rm TMR}^\prime + \hat{g}_{ \rm IZR}^\prime \equiv \hat{g}^\prime$. $\vec{f}_r$ splits accordingly into $\vec{f}_{\rm TMR}+\vec{f}_{ \rm IZR}$, where $\vec{f}_{\rm TMR}$ and $\vec{f}_{ \rm IZR}$ are given by Eq.\,(\ref{eq:Rabi}) using $\hat{g}_{\rm TMR}^\prime$ and $\hat{g}_{ \rm IZR}^\prime$ as input, respectively. $\hat{g}_{\rm TMR}^\prime$ and $\hat{g}_{ \rm IZR}^\prime$ are defined from the decomposition of ${}^{t}\hat{g}\cdot\hat{g}^\prime$ into a symmetric and an antisymmetric matrix. As expected, $\hat{g}_{\rm TMR}^\prime={}^t\hat{g}^{-1}\cdot\hat{G}^\prime/2$ is fully determined by the dependence of the symmetric Zeeman tensor on gate voltage, while $\hat{g}_{ \rm IZR}^\prime$ is totally independent on $\hat{G}^\prime$ (\cite{SupplMat} for details).\\
We compute $\hat{G}^\prime$ from the experimental tensors $\hat{G}(V_0)$ and $\hat{G}(V_1)$ (with $V_1-V_0 = 0.25$\,mV $\simeq V_{ac}$) quoted in Fig.\,\ref{fig:Fig2}a. In the magnetic axes frame $\{\vec{X},\vec{Y},\vec{Z}\}$ at $V_{G2}=V_0$, we find:
\begin{equation}
\hat{G}^\prime(V_0)=
\begin{bmatrix}
-17.9 & 21.1 & 7.2 \\
21.2 & 17.1 & -19.8 \\
7.2 & -19.8 & 9.1
\end{bmatrix}\, \rm V^{-1}.
\label{mat:Zeemanprime}
\end{equation}
We can finally fit the missing $\hat{g}_{ \rm IZR}^\prime$ contribution onto the $f_r$ map of Fig.\,\ref{fig:Fig3}a \cite{SupplMat}. This yields:
\begin{equation}
\hat{g}^\prime(V_0)=
\begin{bmatrix}
-4.3 & -2.4 & -3.2 \\
10.5 & 3.4 & -28.0 \\
8.6 & 30.5 & 2.8
\end{bmatrix}\, \rm V^{-1}
\label{mat:gprime}
\end{equation}
in the same magnetic axes and in the (yet unknown) Kramers basis where $\hat{g}$ is diagonal: $\hat{g}(V_0)={\rm diag}(2.08, 2.48, 1.62)$.\\
Both the Kramers basis and the principal magnetic axes rotate when varying $V_G$ since both $\hat{G}^\prime(V_0)$ and $\hat{g}^\prime(V_0)$ are non-diagonal, which demonstrates concomitant IZ-EDSR and $g$-TMR. The Rabi frequencies calculated from $\hat{g}_d$ and $\hat{g}^\prime$ via Eq.\,(\ref{eq:Rabi}) are plotted in Fig.\,\ref{fig:Fig3}b. We ascribe the small discrepancies to the experimental uncertainty on $\hat{G}^\prime(V_0)$.
\begin{figure}
\centering
	\includegraphics[width=\columnwidth]{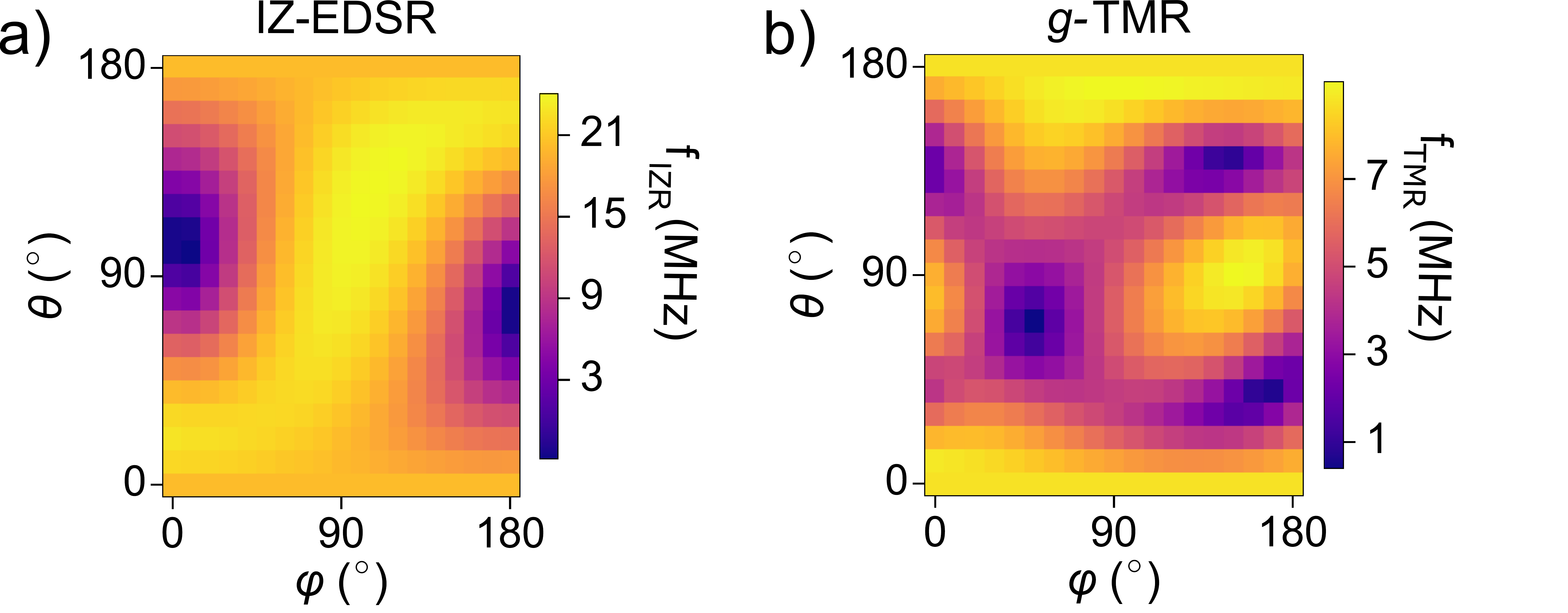} 
	\caption{(Color online) a) IZ-EDSR and b) $g$-TMR contributions to $f_r$. IZ-EDSR is the most relevant contribution, though $g$-TMR can not be neglected as it can reach $\sim30$\% of the experimental Rabi frequencies. Since $|\vec{f}_r| \leq |\vec{f}_{\rm TMR}|+|\vec{f}_{ \rm IZR}|$, the measured values of Fig.\,\ref{fig:Fig3}a are not exactly the sum of  plots a) and b) (see \cite{SupplMat}).}
	\label{fig:Fig4}
\end{figure}
Finally, we address the contributions of IZ-EDSR and $g$-TMR to $f_r$. Figure \ref{fig:Fig4} shows the maps of $f_{ \rm IZR}$ and $f_{\rm TMR}$ obtained from Eq.\,(\ref{eq:Rabi}) using $\hat{g}_{ \rm IZR}^\prime$ and $\hat{g}_{\rm TMR}^\prime$ as inputs, respectively. $f_{\rm TMR}$ (Fig.\,\ref{fig:Fig4}b) is about three times smaller than the maximal $f_{ \rm IZR}$ (Fig.\,\ref{fig:Fig4}a), but still non-negligible. The IZ-EDSR map of Fig.\,\ref{fig:Fig4}a is consistent with an in-plane spin-orbit field $\vec{B}_{\rm SO}\parallel\vec{x}$ (as $f_{ \rm IZR}$ is minimal along that direction). Since holes are mostly confined along $\vec{z}$ by the structure and gates field, the microwave signal on gate 2 hence essentially drives motion of the hole along the nanowire axis $\vec{y}$. As a matter of fact, dot R is controlled by gates 1 and 2 (Fig.\,\ref{fig:Fig1}b), and is likely shifted under the spacer between the gates (around this bias point). A sizable $g$-TMR arises on top of IZ-EDSR due to the changes of the complex confinement potential with gate voltage. Note that the total Rabi frequency $f_r$ also depends on the relative orientation of $\vec{f}_{\rm TMR}$ and $\vec{f}_{ \rm IZR}$ \cite{SupplMat}.\\
\ymn{To conclude, we have shown that spin-orbit-mediated EDSR is generally due to a combination of two effects: the generalized $g$-TMR, essentially associated with a modulation of the confinement potential, and the iso-Zeeman EDSR, which catches, in particular, the motion of the dot wavefunction in the spin-orbit field. While the first one can be determined from the gate-voltage and angular dependence of the Zeeman splitting, the second one can only be extracted from the anisotropy of the Rabi frequency. Both mechanisms can be captured by a linear-in-$B$ model involving a generalized $g$-matrix and its derivative with respect to gate voltage. This model fully characterizes the Larmor and Rabi frequency maps, providing a useful tool for the optimization of spin-orbit qubit devices.}\\
\\
\\
\\
We thank D. Loss, V. N. Golovach and M. Belli for helpful discussions. This work was supported by the European Union's Horizon 2020 research and innovation program under grant agreement No 688539 MOSQUITO and by the ERC project No 759388 LONGSPIN. A.C. and R.M. contributed equally to this work.

\onecolumngrid
\vspace*{1in}

\setcounter{equation}{0}
\setcounter{figure}{0}
\section{SUPPLEMENTAL MATERIAL}

\renewcommand{\theequation}{S\arabic{equation}}  
\renewcommand{\thefigure}{S\arabic{figure}} 

In this supplementary material, we address ($i$) the details of the $g$-matrix formalism for the Rabi frequency, ($ii$) the reformulation in the $g$-matrix formalism of a known example\cite{Golovach} of spin-orbit mediated spin resonance at finite $\vec{B}$, ($iii$) the separation between IZ-EDSR and $g$-TMR contributions, and ($iv$) the experimental procedure to extract IZ-EDSR and $g$-TMR matrices. We also make some concluding remarks on the $g$-matrix formalism. 

\subsection{Model}
\label{sectionModel}

In this Section, we introduce the $g$-matrix and the symmetric Zeeman tensor, and discuss their relations; then we derive the formula for the Rabi frequency in the $g$-matrix formalism.

\subsubsection{The $g$-matrix}

The Hamiltonian of a Kramers doublet $\{\ket{\Uparrow},\ket{\Downarrow}\}$ in a homogeneous magnetic field $\vec{B}$ can be written:
\begin{equation}
H=\frac{1}{2}\mu_B{^t}\vec{\sigma}\cdot\hat{g}\cdot\vec{B}\,,
\label{eqgtensor}
\end{equation}
where $\mu_B$ is Bohr's magneton, $\vec{\sigma}=(\sigma_1,\sigma_2,\sigma_3)$ is the vector of Pauli matrices, $\hat{g}$ is the $g$-matrix (a real $3\times3$ matrix), and $\cdot$ is the matrix product. Any linear-in-$\vec{B}$ two-level Hamiltonian can in principle be mapped onto Eq. (\ref{eqgtensor}) up to an irrelevant energy shift. The 9 elements of the $g$-matrix are independent unless symmetries reduce the number of degrees of freedom. In order to get further insights into the significance of the $g$-matrix, we may factor $\hat{g}=\hat{U}\cdot\hat{g}_d\cdot{^t}\hat{V}$, where $\hat{g}_d={\rm diag}(g_1,g_2,g_3)$ is diagonal and $\hat{U}$, $\hat{V}$ are unitary matrices with determinant $+1$ (singular value decomposition):
\begin{equation}
H=\frac{1}{2}\mu_B{^t}({^t}\hat{U}\cdot\vec{\sigma})\cdot\hat{g}_d\cdot({^t}\hat{V}\cdot\vec{B})\,.
\end{equation}
The columns of $\hat{V}$ define three direct, orthonormal magnetic axes $\vec{X}$, $\vec{Y}$ and $\vec{Z}$. $\vec{\sigma}^\prime={^t}\hat{U}\cdot\vec{\sigma}$ sets three new spin matrices $(\sigma_1^\prime,\sigma_2^\prime,\sigma_3^\prime)$, or, equivalently, three new orthogonal quantization axes for the pseudo-spin of the Kramers doublet. Therefore, there must exist an unitary transform $R$ in the $\{\ket{\Uparrow},\ket{\Downarrow}\}$ subspace such that $R^\dag\sigma_i^\prime R=\sigma_i$ for all $i$'s \cite{noteU}. The columns of $R$ define a new basis $\{\ket{\Uparrow}_{\vec{Z}},\ket{\Downarrow}_{\vec{Z}}\}$ for the two levels system in which:
\begin{equation}
H=\frac{1}{2}\mu_B(g_1B_1\sigma_1+g_2B_2\sigma_2+g_3B_3\sigma_3)\,,
\label{eqgd}
\end{equation}
where $B_1$, $B_2$ and $B_3$ are the components of $\vec{B}$ along the magnetic axes $\vec{X}$, $\vec{Y}$ and $\vec{Z}$. Hence, the $g$-matrix can be made diagonal with an appropriate choice of real space axes for the magnetic field \emph{and} basis set for the two levels system\,\cite{ChibotaruPRL08,Weil-Bolton}. The states $\ket{\Uparrow}_{\vec{Z}}$ and $\ket{\Downarrow}_{\vec{Z}}$ can be identified as the up and down pseudo-spin states along $\vec{Z}$ as they are the eigenstates of $H$ for magnetic fields $\vec{B}\parallel\vec{Z}$.

\subsubsection{The (symmetric) Zeeman tensor}
\label{subsectionZeeman}

Rewriting Eq. (\ref{eqgtensor}) as
\begin{equation}
H=\frac{1}{2}\mu_B|\hat{g}\cdot\vec{B}|\sigma_{\vec{u}}\,,
\end{equation}
where $\sigma_{\vec{u}}={^t}\vec{u}\cdot\vec{\sigma}$ and $\vec{u}=\hat{g}\cdot\vec{B}/|\hat{g}\cdot\vec{B}|$, the Zeeman splitting $\Delta E$ between the eigenstates of $H$ reads:
\begin{equation}
\Delta E=\mu_B|\hat{g}\cdot\vec{B}|\,.
\end{equation}
This can be conveniently cast in the form:
\begin{equation}
\Delta E^2=\mu_B^2({^t}\vec{B}\cdot{^t}\hat{g}\cdot\hat{g}\cdot\vec{B})=\mu_B^2({^t}\vec{B}\cdot\hat{G}\cdot\vec{B})\,,
\end{equation}
where $\hat{G}={^t}\hat{g}\cdot\hat{g}$ is the symmetric Zeeman tensor. From a practical point of view, $\hat{G}$ can be constructed from the measurement of $\Delta E^2$ for six orientations of the magnetic field. Note that $\hat{G}$ only depends on the choice of a frame for the magnetic field. On the contrary $\hat{g}$ depends on the choice of a frame for the magnetic field and on a choice of basis set $\{\ket{\Uparrow},\ket{\Downarrow}\}$ for the Kramers doublet. Any rotation $R$ of the $\{\ket{\Uparrow},\ket{\Downarrow}\}$ basis set results in a corresponding rotation $\hat{g}_R={^t}\hat{U}(R)\cdot\hat{g}$ of the $g$-matrix \cite{noteU}, which leaves the Zeeman tensor $\hat{G}_R={^t}\hat{g}_R\cdot\hat{g}_R={^t}\hat{g}\cdot\hat{g}=\hat{G}$ invariant (as expected, since the Zeeman splittings must not depend on the choice of the $\{\ket{\Uparrow},\ket{\Downarrow}\}$ basis set). It follows from Eq. (\ref{eqgd}) that the eigenvalues of $\hat{G}$ are $g_1^2$, $g_2^2$, and $g_3^2$ while the eigenvectors of $\hat{G}$ are the magnetic axes $\vec{X}$, $\vec{Y}$ and $\vec{Z}$. The characterization of the Zeeman splittings therefore brings the principal $g$ factors and associated magnetic axes, but leaves $\ket{\Uparrow}_{\vec{Z}}$ and $\ket{\Downarrow}_{\vec{Z}}$ unspecified. 

Although not a limitation for many purposes, this may hinder the measurement of $\hat{g}^\prime$, the derivative of $\hat{g}$ with respect to the gate voltage $V$. Indeed, if $\ket{\Uparrow}_{\vec{Z}}$ and $\ket{\Downarrow}_{\vec{Z}}$ depend on $V$, then $\hat{g}(V)$ is diagonal in a different (yet implicit) basis set at each gate voltage. It is then not possible to reconstruct $\hat{g}^\prime$ from the measurement of the Zeeman tensor at different gate voltages.

\subsubsection{The Rabi frequency in the $g$-matrix formalism}

We now derive the formula for the Rabi frequency when the $g$-matrix is dependent on a single control parameter, in this case a gate voltage $V$. When this gate voltage is varied around $V=V_0$, the Hamiltonian can be written:
\begin{align}
H(V)&=\frac{1}{2}\mu_B{^t}\vec{\sigma}\cdot\hat{g}(V)\cdot\vec{B} \nonumber \\
&\simeq\frac{1}{2}\mu_B{^t}\vec{\sigma}\cdot[\hat{g}(V_0)+\hat{g}^\prime(V_0)\delta V]\cdot\vec{B}\,, 
\label{eqgtensorV}
\end{align}
where $\hat{g}^\prime$ is the derivative of $\hat{g}$ with respect to $V$ and $\delta V=V-V_0$. Let us introduce the Larmor vector $\hbar\vec{\Omega}=\mu_B\hat{g}(V_0)\cdot\vec{B}/2$ and its gate-voltage derivative $\hbar\vec{\Omega}^\prime=\mu_B\hat{g}^\prime(V_0)\cdot\vec{B}/2$. Then,
\begin{equation}
H(V)=\hbar|\vec{\Omega}|\sigma_{\vec{\omega}}+\hbar|\vec{\Omega}^\prime|\delta V\sigma_{\vec{\omega}^\prime}\,, \\
\end{equation}
with $\vec{\omega}=\vec{\Omega}/|\vec{\Omega}|$ and $\vec{\omega}^\prime=\vec{\Omega}^\prime/|\vec{\Omega}^\prime|$. Splitting $\vec{\Omega}^\prime=\vec{\Omega}^\prime_\parallel+\vec{\Omega}^\prime_\perp$ into components parallel and perpendicular to $\vec{\Omega}$,
\begin{equation}
H(V)=\hbar|\vec{\Omega}+\vec{\Omega}^\prime_\parallel\delta V|\sigma_{\vec{\omega}}+\hbar|\vec{\Omega}^\prime_\perp|\delta V\sigma_{\vec{\omega}^\prime_\perp}\,. \\
\end{equation}
$\vec{\Omega}^\prime_\parallel$ characterizes gate-driven modulations of the Larmor (spin precession) frequency, while $\vec{\Omega}^\prime_\perp$ characterizes spin rotations. For a radio-frequency (RF) $\delta V=V_{ac}\sin(|\vec{\Omega}|t)$ resonant with the transition between the eigenstates of $H(V_0)$, the Rabi frequency $f_r$ reads\,\cite{Kato,NataliaAPL}:
\begin{align}
hf_r&=\hbar|\vec{\Omega}^\prime_\perp|V_{ac} \nonumber \\
&=\hbar|\vec{\omega}\times\vec{\Omega}^\prime|V_{ac} \nonumber \\
&=\frac{\mu_B B V_{ac}}{2|g^*|}\Big|[\hat{g}(V_0)\cdot\vec{b}]\times[\hat{g}^\prime(V_0)\cdot\vec{b}]\Big|\,,
\label{eqRabi}
\end{align}
where $\vec{b}=\vec{B}/B$ is the unit vector along the magnetic field and $|g^*|=|\hat{g}(V_0)\cdot\vec{b}|$ is the effective $g$ factor along that direction. This may be conveniently written $f_r=|\vec{f}_r|$, with:
\begin{equation}
h\vec{f}_r=\frac{\mu_B B V_{ac}}{2|g^*|}[\hat{g}(V_0)\cdot\vec{b}]\times[\hat{g}^\prime(V_0)\cdot\vec{b}]\,.
\label{eqfr}
\end{equation}
Note that the Larmor frequency $|\vec{\Omega}|/(2\pi)$ and the Rabi frequency $|\vec{f_r}|$ do not depend on the choice of the basis set for the Kramers doublet. The orientation of $\vec{\Omega}$ and $\vec{f_r}$, however, does (any change of $\{\ket{\Uparrow},\ket{\Downarrow}\}$ basis set results in a rotation of $\vec{\Omega}$ and $\vec{f_r}$) \cite{noteU}. $\vec{\Omega}$ and $\vec{f_r}$ actually define the axis of precession and the axis of rotation of the pseudo-spin in the Bloch sphere defined by the $\{\ket{\Uparrow},\ket{\Downarrow}\}$ states.

\subsection{Example: Electric Dipole Spin Resonance in a finite magnetic field}
\label{sectionExample}

Any mechanism giving rise to Rabi oscillations with frequency proportional to $B$ and $V_{ac}$ shall be captured by Eq. (\ref{eqRabi}). This excludes spin rotations at zero field \cite{Golovach10} or showing significant non-linearities.\cite{Corna17} 

As an example, we can cast the spin-orbit mechanism of Ref. \onlinecite{Golovach} in the $g$-matrix formalism. We consider a quantum dot in the effective mass approximation, with strong confinement along $\boldsymbol{z}$, harmonic confinement $V(x,y)=m\omega_0^2(x^2+y^2)/2$ in the $(xy)$ plane, and in-plane Rashba plus Dresselhaus spin-orbit coupling.\cite{Golovach} We assume an isotropic $g$ factor $g_0$. An electric field $\vec{E}=E_0(e_x\vec{x}+e_y\vec{y})\sin(\omega t)$ is applied in the $(xy)$ plane in order to drive Rabi oscillations between the $\ket{\downarrow}$ and $\ket{\uparrow}$ states ($e_x^2+e_y^2=1$, the spin being quantized along $\boldsymbol{z}$). Then, according to Ref. \onlinecite{Golovach},
\begin{equation}
hf_r=2g_0\mu_B|\vec{B}\times\vec{\Omega}_0|\,,
\label{eqRabiLoss}
\end{equation}
at resonance, where:
\begin{equation}
\vec{\Omega}_0=-\frac{eE_0}{m\omega_0^2}\left(\frac{e_y}{\lambda_-}, \frac{e_x}{\lambda_+}, 0\right)\,,
\end{equation}
and $\lambda_\pm=\hbar/[m(\beta\pm\alpha)]$, ($\alpha$, $\beta$) being the Rashba and Dresselhaus spin-orbit constants. From the effective Hamiltonian of the quantum dot in the static limit ($\omega\to 0$),\cite{Golovach}
\begin{equation}
H_{\rm eff}=\frac{1}{2}g_0\mu_B\vec{B}\cdot\vec{\sigma}+g_0\mu_B(\vec{B}\times\vec{\Omega}_0)\cdot\vec{\sigma}\,,
\end{equation}
the matrices $\hat{g}$ and $\hat{g}^\prime$ can be readily identified (the derivative being taken with respect to $E_0$ instead of $V$):
\begin{subequations}
\label{eqggprime}
\begin{align}
\hat{g}&=g_0\begin{pmatrix}{}
1 & 0 & 0 \\
0 & 1 & 0 \\
0 & 0 & 1
\end{pmatrix} \\
\hat{g}^\prime&=2g_0
\begin{pmatrix}{}
0 & 0 & -\Theta_{0,y} \\
0 & 0 & +\Theta_{0,x} \\
+\Theta_{0,y} & -\Theta_{0,x} & 0
\end{pmatrix}\,,
\end{align}
\end{subequations}
with $\vec{\Theta}_0=\vec{\Omega}_0/E_0$. Insertion of Eqs. (\ref{eqggprime}) into Eqs. (\ref{eqRabi}) yields back Eq. (\ref{eqRabiLoss}), showing that the present $g$-matrix formalism indeed captures the EDSR mechanism of Ref. \onlinecite{Golovach}.

\subsection{Discussion: IZ-EDSR vs $g$-TMR}
\label{sectionIZ-EDSR$g$-TMR}

In the above set-up the Rabi oscillations are mediated by spin-orbit coupling when the quantum dot is moved around its equilibrium position by the alternating electric field. This example highlights the distinction between the IZ-EDSR and $g$-TMR contributions introduced in the main text.

Indeed, in a harmonic potential, the alternating electric field moves the center of the dot around but does not change the harmonic shape of the confinement potential. Therefore, the Zeeman splittings are expected to be independent on the position of the dot. As a matter of fact,
\begin{equation}
\hat{G}^\prime={}^{t}\hat{g}\cdot\hat{g}^\prime+{}^t\hat{g}^\prime\cdot\hat{g}
\label{eqgprime}
\end{equation}
is zero because $\hat{g}^\prime$ is antisymmetric. While gauge invariance imposes that $\hat{G}$ does not change upon translation of the dot in an uniform magnetic field, $\hat{g}$ does depend on the position of the dot because the magnetic vector potential breaks translational symmetry. 

The set-up of Ref. \onlinecite{Golovach} hence gives rise to pure ``IZ-EDSR'' (i.e., modulations of the $g$-matrix without modulations of the Zeeman splittings). In an arbitrary potential/non-uniform alternating electric field, the motion of the dot will, in general, come along with changes of the shape of the confinement potential resulting in modulations of the Zeeman splittings (extra $g$-TMR contributions to $\hat{g}^\prime$). These modulations of the Zeeman splittings share, however, the same microscopic origin as IZ-EDSR, namely spin-orbit coupling, and shall therefore be treated on the same footing.

As discussed in the main text, we can always split ${}^{t}\hat{g}\cdot\hat{g}^\prime=\hat{S}+\hat{A}$, where $\hat{S}$ is a symmetric and $\hat{A}$ is an antisymmetric matrix. Then, Eq. (\ref{eqgprime}) sets $\hat{S}=\hat{G}^\prime/2$. We can next introduce the $g$-TMR matrix 
\begin{equation}
\hat{g}_{\rm TMR}^\prime={}^t\hat{g}^{-1}\cdot\hat{G}^\prime/2
\end{equation}
and the IZ-EDSR matrix 
\begin{equation}
\hat{g}_{\rm IZR}^\prime={}^t\hat{g}^{-1}\cdot\hat{A},
\end{equation}
and split Eq. (\ref{eqfr}) as $\vec{f}_r=\vec{f}_{\rm TMR}+\vec{f}_{\rm IZR}$, where $\vec{f}_{\rm TMR}$ and $\vec{f}_{\rm IZR}$ are the contributions of $\hat{g}_{\rm TMR}^\prime$ and $\hat{g}_{\rm IZR}^\prime$ to the Rabi frequency. The $g$-TMR matrix captures the modulations of the Zeeman tensor, while the IZ-EDSR matrix captures the pure rotations of $\{\ket{\Uparrow}_{\vec{Z}},\ket{\Downarrow}_{\vec{Z}}\}$ that do not give rise to modulations of the Zeeman splittings. If the Rabi oscillations are mediated by spin-orbit coupling,\cite{Golovach} $g$-TMR typically results from the changes of the shape of the potential, while IZ-EDSR also captures the contribution from the motion of the dot in the electric field.

In most situations IZ-EDSR coexists with $g$-TMR. IZ-EDSR is expected to dominate over $g$-TMR whenever the motion of the dot does not come along with significant variations of the shape of the potential, as outlined in Ref. \onlinecite{Golovach}. On the opposite, $g$-TMR prevails in highly symmetric situations and/or for specific orientations of the magnetic field where the gate potential only modulates the $g_i^*$'s (as in Ref. \onlinecite{NataliaAPL}). If the gate potential also controls the magnetic axes $\{\vec{X},\vec{Y},\vec{Z}\}$, $g$-TMR is typically accompanied by IZ-EDSR.

\subsection{Extraction of IZ-EDSR and $g$-TMR contributions}
\label{sectionexp}

As discussed in section \ref{subsectionZeeman}, the $\hat{g}^\prime$ matrix can not usually be reconstructed from the measurement of the Zeeman splittings since the IZ-EDSR contribution does not give rise to modulations of the Zeeman tensor. Nonetheless, $\hat{g}_{\rm IZR}^\prime$ can be extracted from the Rabi frequency map. We detail the procedure below. 

First, the symmetric Zeeman tensor $\hat{G}={^t}\hat{g}\cdot\hat{g}$ is constructed from the measurement of the Zeeman splittings along 6 independent directions. The eigenvalues of $\hat{G}$ are the square of the principal $g$ factors $g_1^2$, $g_2^2$, and $g_3^2$, while the eigenvectors of $\hat{G}$ are the principal magnetic axes $\vec{X}$, $\vec{Y}$ and $\vec{Z}$. In the magnetic axes frame, there exists a (yet implicit) basis set $\{\ket{\Uparrow}_{\vec{Z}},\ket{\Downarrow}_{\vec{Z}}\}$ for the Kramers doublet such that:
\begin{equation}
\hat{g}\equiv\hat{g}_d=
\begin{pmatrix}{}
g_1 & 0 & 0 \\
0 & g_2 & 0 \\
0 & 0 & g_3
\end{pmatrix}\,.
\end{equation}
Note that there might be an ambiguity on the sign of the $g_i$'s (all assumed with the same sign here).

The matrix $\hat{G}^\prime$ is extracted from the measurement of $\hat{G}$ for two nearby gate voltages $V=V_0$ and $V=V_0+\delta V$. In the principal magnetic axes and $\{\ket{\Uparrow}_{\vec{Z}},\ket{\Downarrow}_{\vec{Z}}\}$ basis set at $V=V_0$, the $g$-TMR matrix is then $\hat{g}_{\rm TMR}^\prime=\hat{g}_d^{-1}\cdot\hat{G}^\prime/2$.

The three independent elements of $\hat{A}$ are fitted to the dependence of the Rabi frequency on the magnetic field orientation using Eq. (\ref{eqRabi}) (in principle, a measurement of the Rabi frequency along three directions shall be sufficient; here we made a least square fit on the whole map). The amplitude $V_{ac}$ of the RF field needs to be included in the fit if the attenuation of the RF lines is not exactly known. Note that the Rabi frequency is measured at constant Zeeman splitting (resonance frequency $|\vec{\Omega}|/(2\pi)=9$ GHz) instead of constant magnetic field in order to ensure that the attenuation does not vary from one field orientation to an other.

This procedure yields in the present case $g_1(V_0)=2.08$ along $\vec{X}=(0.82,0.19,-0.53)$, $g_2(V_0)=2.48$ along $\vec{Y}=(-0.22,0.98,0.01)$, $g_3(V_0)=1.62$ along $\vec{Z}=(0.52,0.11,0.84)$, 
\begin{equation}
\hat{g}^\prime_{\rm TMR}(V_0)=
\begin{pmatrix}{}
-4.31 &  5.07 &  1.73 \\
 4.26 &  3.45 & -4.01 \\
 2.22 & -6.12 &  2.82
\end{pmatrix}{\rm V}^{-1}\,,
\end{equation}
and
\begin{equation}
\hat{g}^\prime_{\rm IZR}(V_0)=
\begin{pmatrix}{}
0.0 & -7.45 & -4.97 \\
6.26 & 0.0 & -23.99 \\
6.38 & 36.60 & 0.0
\end{pmatrix}{\rm V}^{-1}\,.
\end{equation}
We estimate $V_{ac}=0.41$ mV, close to the value expected for the RF setup. The fact that ${G}^\prime$ is not diagonal in this low-symmetry device shows that the principal magnetic axes (as well as, presumably, the basis set $\{\ket{\Uparrow}_{\vec{Z}},\ket{\Downarrow}_{\vec{Z}}\}$) rotate with the gate voltage. The calculated map of Rabi frequencies $f_r$ is plotted on Fig.\,3b of the main text, while the IZ-EDSR contribution $|\vec{f}_{\rm IZR}|$ and the $g$-TMR contribution $|\vec{f}_{\rm TMR}|$ are plotted in Fig.\,4a and Fig.\,4b, respectively. 

The $g$-TMR map of Fig.\,4b shows a complex dependence on the magnetic field orientation that reflects the complex structure of the potential in the device. As for the IZ-EDSR diagram of Fig.\,4a, we point out that, although the model of Ref.\,\onlinecite{Golovach} can not be applied directly, $|\vec{f}_{\rm IZR}|$ reproduces one of the most salient feature of Eq. (\ref{eqRabiLoss}): two well-defined minima along $\vec{x}=[1\bar{1}0]$ suggest that the effective spin-orbit field is perpendicular to the nanowire. Since the static electric field in the dot is dominated by the vertical ($\parallel\vec{z}$) component, this implies that the RF electric field on gate 2 essentially drives motion of the hole in dot R along the nanowire axis $\vec{y}$. This is consistent with the fact that dot R is controlled by both gates 1 and 2 on Fig. 1b of the main text, and is, therefore, likely located under the spacer between gate 1 and gate 2 at this bias point. This calls for a detailed modeling of the potential and spin-orbit interactions in these devices, however beyond the scope of this paper.

\begin{figure}
\centering
\includegraphics[width=.9\columnwidth]{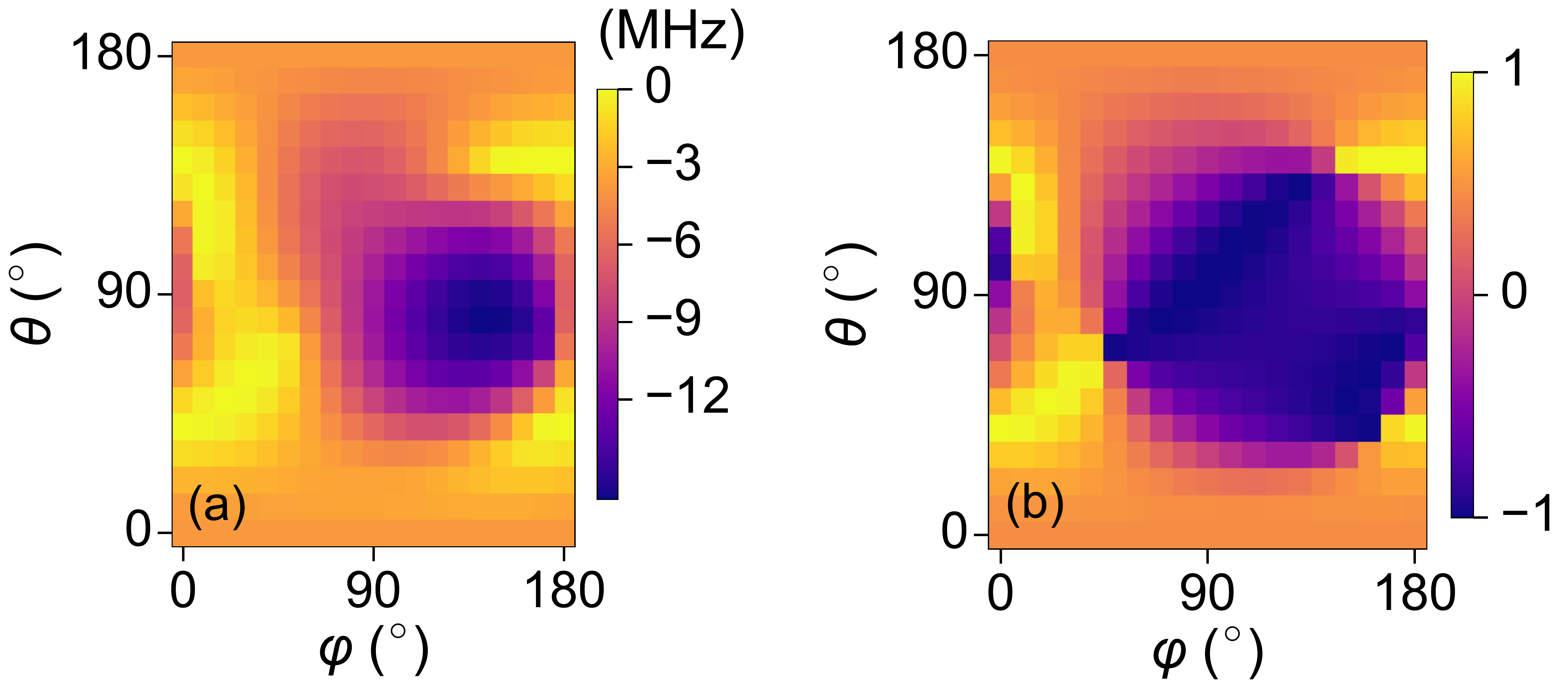} 
\caption{(a) Map of $|\vec{f}_r|-|\vec{f}_{\rm IZR}|-|\vec{f}_{\rm TMR}|$ as a function of the orientation of the magnetic field (see main text for a definition of $\theta$ and $\phi$. (b) Map of $\cos(\theta_Z)$ as a function of the orientation of the magnetic field, where $\theta_Z$ is the angle between $\vec{f}_{\rm IZR}$ and $\vec{f}_{\rm TMR}$. These two maps are plotted at constant Zeeman splitting $|g^*|\mu_BB/h=9$ GHz as in the main text.}
\label{figS1}
\end{figure}

We emphasize that, in general, $f_r<|\vec{f}_{\rm IZR}|+|\vec{f}_{\rm TMR}|$, because $\vec{f}_{\rm IZR}$ and $\vec{f}_{\rm TMR}$ are not aligned (IZ-EDSR and $g$-TMR drive rotations around different axes). In order to highlight this correction, we have plotted in Fig. \ref{figS1} the difference $f_r-|\vec{f}_{\rm IZR}|-|\vec{f}_{\rm TMR}|$ along with $\cos(\theta_Z)$, where $\theta_Z$ is the angle between $\vec{f}_{\rm IZR}$ and $\vec{f}_{\rm TMR}$. There is a large sector around $\theta=90^\circ$, $\varphi=135^\circ$ where IZ-EDSR and $g$-TMR tend to cancel each other. The design of the device might be optimized in order to align $\vec{f}_{\rm IZR}$ and $\vec{f}_{\rm TMR}$ as best as possible.

\subsection{Concluding remarks}
\label{sectionconclusions}

As discussed earlier, Eq. (\ref{eqRabi}) holds in any device where the Rabi frequencies at finite magnetic field are linear in $B$ and $V_{ac}$. It can therefore be used as a ``compact model'' for the control of the device. 
The analysis of the orientational dependence of $\vec{f}_r$, $\vec{f}_{\rm IZR}$ and $\vec{f}_{\rm TMR}$ combined with appropriate modeling may provide valuable information about the confinement potential and spin-orbit interactions in the qubits.

As far as modeling is concerned, the $g$-matrix formalism is a very efficient way to compute the map of Rabi frequencies. Indeed, the $g$-matrix at a given gate voltage $V=V_0$ can be calculated from the eigenstates at zero magnetic field using simple perturbation theory \cite{Ivchenko,Tadjine17} and its derivative can be obtained from finite differences between gate voltages $V=V_0$ and $V=V_0+\delta V$ (note that in a numerical calculation the basis set $\{\ket{\Uparrow},\ket{\Downarrow}\}$ can be set explicitly). The complete map of Rabi frequencies can hence be constructed from two electronic structure calculations at zero magnetic field but different gate voltages, without the need for an explicit sampling over magnetic field orientations.

\twocolumngrid


\begin{thebibliography}{37}%
\makeatletter
\providecommand \@ifxundefined [1]{%
 \@ifx{#1\undefined}
}%
\providecommand \@ifnum [1]{%
 \ifnum #1\expandafter \@firstoftwo
 \else \expandafter \@secondoftwo
 \fi
}%
\providecommand \@ifx [1]{%
 \ifx #1\expandafter \@firstoftwo
 \else \expandafter \@secondoftwo
 \fi
}%
\providecommand \natexlab [1]{#1}%
\providecommand \enquote  [1]{``#1''}%
\providecommand \bibnamefont  [1]{#1}%
\providecommand \bibfnamefont [1]{#1}%
\providecommand \citenamefont [1]{#1}%
\providecommand \href@noop [0]{\@secondoftwo}%
\providecommand \href [0]{\begingroup \@sanitize@url \@href}%
\providecommand \@href[1]{\@@startlink{#1}\@@href}%
\providecommand \@@href[1]{\endgroup#1\@@endlink}%
\providecommand \@sanitize@url [0]{\catcode `\\12\catcode `\$12\catcode
  `\&12\catcode `\#12\catcode `\^12\catcode `\_12\catcode `\%12\relax}%
\providecommand \@@startlink[1]{}%
\providecommand \@@endlink[0]{}%
\providecommand \url  [0]{\begingroup\@sanitize@url \@url }%
\providecommand \@url [1]{\endgroup\@href {#1}{\urlprefix }}%
\providecommand \urlprefix  [0]{URL }%
\providecommand \Eprint [0]{\href }%
\providecommand \doibase [0]{http://dx.doi.org/}%
\providecommand \selectlanguage [0]{\@gobble}%
\providecommand \bibinfo  [0]{\@secondoftwo}%
\providecommand \bibfield  [0]{\@secondoftwo}%
\providecommand \translation [1]{[#1]}%
\providecommand \BibitemOpen [0]{}%
\providecommand \bibitemStop [0]{}%
\providecommand \bibitemNoStop [0]{.\EOS\space}%
\providecommand \EOS [0]{\spacefactor3000\relax}%
\providecommand \BibitemShut  [1]{\csname bibitem#1\endcsname}%
\let\auto@bib@innerbib\@empty
\bibitem [{\citenamefont {Rashba}\ and\ \citenamefont
  {Sheka}(1991)}]{RashbaBook}%
  \BibitemOpen
  \bibfield  {author} {\bibinfo {author} {\bibfnamefont {E.~I.}\ \bibnamefont
  {Rashba}}\ and\ \bibinfo {author} {\bibfnamefont {V.~I.}\ \bibnamefont
  {Sheka}},\ }\href@noop {} {\emph {\bibinfo {title} {Landau Level
  Spectroscopy}}}\ (\bibinfo  {publisher} {North-Holland, Amsterdam},\ \bibinfo
  {year} {1991})\BibitemShut {NoStop}%
\bibitem [{\citenamefont {Rashba}(2008)}]{rashbaReview}%
  \BibitemOpen
  \bibfield  {author} {\bibinfo {author} {\bibfnamefont {E.~I.}\ \bibnamefont
  {Rashba}},\ }\href@noop {} {\bibfield  {journal} {\bibinfo  {journal} {Phys.
  Rev. B}\ }\textbf {\bibinfo {volume} {78}},\ \bibinfo {pages} {195302}
  (\bibinfo {year} {2008})}\BibitemShut {NoStop}%
\bibitem [{\citenamefont {Koppens}\ \emph {et~al.}(2006)\citenamefont
  {Koppens}, \citenamefont {Buizert}, \citenamefont {Tielrooij}, \citenamefont
  {Vink}, \citenamefont {Nowack}, \citenamefont {Meunier}, \citenamefont
  {Kouwenhoven},\ and\ \citenamefont {Vandersypen}}]{Koppens}%
  \BibitemOpen
  \bibfield  {author} {\bibinfo {author} {\bibfnamefont {F.}~\bibnamefont
  {Koppens}}, \bibinfo {author} {\bibfnamefont {C.}~\bibnamefont {Buizert}},
  \bibinfo {author} {\bibfnamefont {K.}~\bibnamefont {Tielrooij}}, \bibinfo
  {author} {\bibfnamefont {I.}~\bibnamefont {Vink}}, \bibinfo {author}
  {\bibfnamefont {K.}~\bibnamefont {Nowack}}, \bibinfo {author} {\bibfnamefont
  {T.}~\bibnamefont {Meunier}}, \bibinfo {author} {\bibfnamefont
  {L.}~\bibnamefont {Kouwenhoven}}, \ and\ \bibinfo {author} {\bibfnamefont
  {L.}~\bibnamefont {Vandersypen}},\ }\href@noop {} {\bibfield  {journal}
  {\bibinfo  {journal} {Nature}\ }\textbf {\bibinfo {volume} {442}},\ \bibinfo
  {pages} {766} (\bibinfo {year} {2006})}\BibitemShut {NoStop}%
\bibitem [{\citenamefont {Pla}\ \emph {et~al.}(2012)\citenamefont {Pla},
  \citenamefont {Tan}, \citenamefont {Dehollain}, \citenamefont {Lim},
  \citenamefont {Morton}, \citenamefont {Jamieson}, \citenamefont {Dzurak},\
  and\ \citenamefont {Morello}}]{pla_electronqubit}%
  \BibitemOpen
  \bibfield  {author} {\bibinfo {author} {\bibfnamefont {J.~J.}\ \bibnamefont
  {Pla}}, \bibinfo {author} {\bibfnamefont {K.~Y.}\ \bibnamefont {Tan}},
  \bibinfo {author} {\bibfnamefont {J.~P.}\ \bibnamefont {Dehollain}}, \bibinfo
  {author} {\bibfnamefont {W.~H.}\ \bibnamefont {Lim}}, \bibinfo {author}
  {\bibfnamefont {J.~J.}\ \bibnamefont {Morton}}, \bibinfo {author}
  {\bibfnamefont {D.~N.}\ \bibnamefont {Jamieson}}, \bibinfo {author}
  {\bibfnamefont {A.~S.}\ \bibnamefont {Dzurak}}, \ and\ \bibinfo {author}
  {\bibfnamefont {A.}~\bibnamefont {Morello}},\ }\href@noop {} {\bibfield
  {journal} {\bibinfo  {journal} {Nature}\ }\textbf {\bibinfo {volume} {489}},\
  \bibinfo {pages} {541} (\bibinfo {year} {2012})}\BibitemShut {NoStop}%
\bibitem [{\citenamefont {Veldhorst}\ \emph {et~al.}(2014)\citenamefont
  {Veldhorst}, \citenamefont {Hwang}, \citenamefont {Yang}, \citenamefont
  {Leenstra}, \citenamefont {De~Ronde}, \citenamefont {Dehollain},
  \citenamefont {Muhonen}, \citenamefont {Hudson}, \citenamefont {Itoh},
  \citenamefont {Morello},\ and\ \citenamefont
  {Dzurak}}]{VeldhorstAddressable}%
  \BibitemOpen
  \bibfield  {author} {\bibinfo {author} {\bibfnamefont {M.}~\bibnamefont
  {Veldhorst}}, \bibinfo {author} {\bibfnamefont {J.~C.~C.}\ \bibnamefont
  {Hwang}}, \bibinfo {author} {\bibfnamefont {C.~H.}\ \bibnamefont {Yang}},
  \bibinfo {author} {\bibfnamefont {A.~W.}\ \bibnamefont {Leenstra}}, \bibinfo
  {author} {\bibfnamefont {B.}~\bibnamefont {De~Ronde}}, \bibinfo {author}
  {\bibfnamefont {J.~P.}\ \bibnamefont {Dehollain}}, \bibinfo {author}
  {\bibfnamefont {J.~T.}\ \bibnamefont {Muhonen}}, \bibinfo {author}
  {\bibfnamefont {F.~E.}\ \bibnamefont {Hudson}}, \bibinfo {author}
  {\bibfnamefont {K.~M.}\ \bibnamefont {Itoh}}, \bibinfo {author}
  {\bibfnamefont {A.}~\bibnamefont {Morello}}, \ and\ \bibinfo {author}
  {\bibfnamefont {A.~S.}\ \bibnamefont {Dzurak}},\ }\href@noop {} {\bibfield
  {journal} {\bibinfo  {journal} {Nature Nanotech.}\ }\textbf {\bibinfo
  {volume} {9}},\ \bibinfo {pages} {981} (\bibinfo {year} {2014})}\BibitemShut
  {NoStop}%
\bibitem [{\citenamefont {Yoneda}\ \emph {et~al.}(2014)\citenamefont {Yoneda},
  \citenamefont {Otsuka}, \citenamefont {Nakajima}, \citenamefont {Takakura},
  \citenamefont {Obata}, \citenamefont {Pioro-Ladri{\'e}re}, \citenamefont
  {Lu}, \citenamefont {Palmstr{\o}m}, \citenamefont {Gossard},\ and\
  \citenamefont {Tarucha}}]{YonedaPRL2014}%
  \BibitemOpen
  \bibfield  {author} {\bibinfo {author} {\bibfnamefont {J.}~\bibnamefont
  {Yoneda}}, \bibinfo {author} {\bibfnamefont {T.}~\bibnamefont {Otsuka}},
  \bibinfo {author} {\bibfnamefont {T.}~\bibnamefont {Nakajima}}, \bibinfo
  {author} {\bibfnamefont {T.}~\bibnamefont {Takakura}}, \bibinfo {author}
  {\bibfnamefont {T.}~\bibnamefont {Obata}}, \bibinfo {author} {\bibfnamefont
  {M.}~\bibnamefont {Pioro-Ladri{\'e}re}}, \bibinfo {author} {\bibfnamefont
  {H.}~\bibnamefont {Lu}}, \bibinfo {author} {\bibfnamefont {C.}~\bibnamefont
  {Palmstr{\o}m}}, \bibinfo {author} {\bibfnamefont {A.}~\bibnamefont
  {Gossard}}, \ and\ \bibinfo {author} {\bibfnamefont {S.}~\bibnamefont
  {Tarucha}},\ }\href@noop {} {\bibfield  {journal} {\bibinfo  {journal} {Phys.
  Rev. Lett.}\ }\textbf {\bibinfo {volume} {113}},\ \bibinfo {pages} {267601}
  (\bibinfo {year} {2014})}\BibitemShut {NoStop}%
\bibitem [{\citenamefont {Maurand}\ \emph {et~al.}(2016)\citenamefont
  {Maurand}, \citenamefont {Jehl}, \citenamefont {Kotekar-Patil}, \citenamefont
  {Corna}, \citenamefont {Bohuslavskyi}, \citenamefont {Lavi{\'e}ville},
  \citenamefont {Hutin}, \citenamefont {Barraud}, \citenamefont {Vinet},
  \citenamefont {Sanquer},\ and\ \citenamefont {De~Franceschi}}]{roro}%
  \BibitemOpen
  \bibfield  {author} {\bibinfo {author} {\bibfnamefont {R.}~\bibnamefont
  {Maurand}}, \bibinfo {author} {\bibfnamefont {X.}~\bibnamefont {Jehl}},
  \bibinfo {author} {\bibfnamefont {D.}~\bibnamefont {Kotekar-Patil}}, \bibinfo
  {author} {\bibfnamefont {A.}~\bibnamefont {Corna}}, \bibinfo {author}
  {\bibfnamefont {H.}~\bibnamefont {Bohuslavskyi}}, \bibinfo {author}
  {\bibfnamefont {R.}~\bibnamefont {Lavi{\'e}ville}}, \bibinfo {author}
  {\bibfnamefont {L.}~\bibnamefont {Hutin}}, \bibinfo {author} {\bibfnamefont
  {S.}~\bibnamefont {Barraud}}, \bibinfo {author} {\bibfnamefont
  {M.}~\bibnamefont {Vinet}}, \bibinfo {author} {\bibfnamefont
  {M.}~\bibnamefont {Sanquer}}, \ and\ \bibinfo {author} {\bibfnamefont
  {S.}~\bibnamefont {De~Franceschi}},\ }\href@noop {} {\bibfield  {journal}
  {\bibinfo  {journal} {Nature Comm.}\ }\textbf {\bibinfo {volume} {7}},\
  \bibinfo {pages} {13575} (\bibinfo {year} {2016})}\BibitemShut {NoStop}%
\bibitem [{\citenamefont {Yoneda}\ \emph {et~al.}(2018)\citenamefont {Yoneda},
  \citenamefont {Takeda}, \citenamefont {Otsuka}, \citenamefont {Nakajima},
  \citenamefont {Delbecq}, \citenamefont {Allison}, \citenamefont {Honda},
  \citenamefont {Kodera}, \citenamefont {Oda}, \citenamefont {Hoshi},
  \citenamefont {Usami}, \citenamefont {Itoh},\ and\ \citenamefont
  {Tarucha}}]{Yoneda2017}%
  \BibitemOpen
  \bibfield  {author} {\bibinfo {author} {\bibfnamefont {J.}~\bibnamefont
  {Yoneda}}, \bibinfo {author} {\bibfnamefont {K.}~\bibnamefont {Takeda}},
  \bibinfo {author} {\bibfnamefont {T.}~\bibnamefont {Otsuka}}, \bibinfo
  {author} {\bibfnamefont {T.}~\bibnamefont {Nakajima}}, \bibinfo {author}
  {\bibfnamefont {M.}~\bibnamefont {Delbecq}}, \bibinfo {author} {\bibfnamefont
  {G.}~\bibnamefont {Allison}}, \bibinfo {author} {\bibfnamefont
  {T.}~\bibnamefont {Honda}}, \bibinfo {author} {\bibfnamefont
  {T.}~\bibnamefont {Kodera}}, \bibinfo {author} {\bibfnamefont
  {S.}~\bibnamefont {Oda}}, \bibinfo {author} {\bibfnamefont {Y.}~\bibnamefont
  {Hoshi}}, \bibinfo {author} {\bibfnamefont {N.}~\bibnamefont {Usami}},
  \bibinfo {author} {\bibfnamefont {K.}~\bibnamefont {Itoh}}, \ and\ \bibinfo
  {author} {\bibfnamefont {S.}~\bibnamefont {Tarucha}},\ }\href@noop {}
  {\bibfield  {journal} {\bibinfo  {journal} {Nat. Nanotech.}\ }\textbf
  {\bibinfo {volume} {13}},\ \bibinfo {pages} {102} (\bibinfo {year}
  {2018})}\BibitemShut {NoStop}%
\bibitem [{\citenamefont {Kato}\ \emph {et~al.}(2003)\citenamefont {Kato},
  \citenamefont {Myers}, \citenamefont {Driscoll}, \citenamefont {Gossard},
  \citenamefont {Levy},\ and\ \citenamefont {Awschalom}}]{Kato}%
  \BibitemOpen
  \bibfield  {author} {\bibinfo {author} {\bibfnamefont {Y.}~\bibnamefont
  {Kato}}, \bibinfo {author} {\bibfnamefont {R.}~\bibnamefont {Myers}},
  \bibinfo {author} {\bibfnamefont {D.}~\bibnamefont {Driscoll}}, \bibinfo
  {author} {\bibfnamefont {A.}~\bibnamefont {Gossard}}, \bibinfo {author}
  {\bibfnamefont {J.}~\bibnamefont {Levy}}, \ and\ \bibinfo {author}
  {\bibfnamefont {D.}~\bibnamefont {Awschalom}},\ }\href@noop {} {\bibfield
  {journal} {\bibinfo  {journal} {Science}\ }\textbf {\bibinfo {volume}
  {299}},\ \bibinfo {pages} {1201} (\bibinfo {year} {2003})}\BibitemShut
  {NoStop}%
\bibitem [{\citenamefont {Pingenot}\ \emph {et~al.}(2011)\citenamefont
  {Pingenot}, \citenamefont {Pryor},\ and\ \citenamefont
  {Flatt{\'e}}}]{FlattePRB}%
  \BibitemOpen
  \bibfield  {author} {\bibinfo {author} {\bibfnamefont {J.}~\bibnamefont
  {Pingenot}}, \bibinfo {author} {\bibfnamefont {C.~E.}\ \bibnamefont {Pryor}},
  \ and\ \bibinfo {author} {\bibfnamefont {M.~E.}\ \bibnamefont {Flatt{\'e}}},\
  }\href@noop {} {\bibfield  {journal} {\bibinfo  {journal} {Phys. Rev. B}\
  }\textbf {\bibinfo {volume} {84}},\ \bibinfo {pages} {195403} (\bibinfo
  {year} {2011})}\BibitemShut {NoStop}%
\bibitem [{\citenamefont {Schroer}\ \emph {et~al.}(2011)\citenamefont
  {Schroer}, \citenamefont {Petersson}, \citenamefont {Jung},\ and\
  \citenamefont {Petta}}]{PettaGtuning}%
  \BibitemOpen
  \bibfield  {author} {\bibinfo {author} {\bibfnamefont {M.}~\bibnamefont
  {Schroer}}, \bibinfo {author} {\bibfnamefont {K.}~\bibnamefont {Petersson}},
  \bibinfo {author} {\bibfnamefont {M.}~\bibnamefont {Jung}}, \ and\ \bibinfo
  {author} {\bibfnamefont {J.}~\bibnamefont {Petta}},\ }\href@noop {}
  {\bibfield  {journal} {\bibinfo  {journal} {Phys. Rev. Lett.}\ }\textbf
  {\bibinfo {volume} {107}},\ \bibinfo {pages} {176811} (\bibinfo {year}
  {2011})}\BibitemShut {NoStop}%
\bibitem [{\citenamefont {Takahashi}\ \emph {et~al.}(2013)\citenamefont
  {Takahashi}, \citenamefont {Deacon}, \citenamefont {Oiwa}, \citenamefont
  {Shibata}, \citenamefont {Hirakawa},\ and\ \citenamefont
  {Tarucha}}]{TaruchaPRBR13}%
  \BibitemOpen
  \bibfield  {author} {\bibinfo {author} {\bibfnamefont {S.}~\bibnamefont
  {Takahashi}}, \bibinfo {author} {\bibfnamefont {R.}~\bibnamefont {Deacon}},
  \bibinfo {author} {\bibfnamefont {A.}~\bibnamefont {Oiwa}}, \bibinfo {author}
  {\bibfnamefont {K.}~\bibnamefont {Shibata}}, \bibinfo {author} {\bibfnamefont
  {K.}~\bibnamefont {Hirakawa}}, \ and\ \bibinfo {author} {\bibfnamefont
  {S.}~\bibnamefont {Tarucha}},\ }\href@noop {} {\bibfield  {journal} {\bibinfo
   {journal} {Phys. Rev. B}\ }\textbf {\bibinfo {volume} {87}},\ \bibinfo
  {pages} {161302} (\bibinfo {year} {2013})}\BibitemShut {NoStop}%
\bibitem [{\citenamefont {Ares}\ \emph {et~al.}(2013)\citenamefont {Ares},
  \citenamefont {Katsaros}, \citenamefont {Golovach}, \citenamefont {Zhang},
  \citenamefont {Prager}, \citenamefont {Glazman}, \citenamefont {Schmidt},\
  and\ \citenamefont {De~Franceschi}}]{NataliaAPL}%
  \BibitemOpen
  \bibfield  {author} {\bibinfo {author} {\bibfnamefont {N.}~\bibnamefont
  {Ares}}, \bibinfo {author} {\bibfnamefont {G.}~\bibnamefont {Katsaros}},
  \bibinfo {author} {\bibfnamefont {V.}~\bibnamefont {Golovach}}, \bibinfo
  {author} {\bibfnamefont {J.}~\bibnamefont {Zhang}}, \bibinfo {author}
  {\bibfnamefont {A.}~\bibnamefont {Prager}}, \bibinfo {author} {\bibfnamefont
  {L.}~\bibnamefont {Glazman}}, \bibinfo {author} {\bibfnamefont
  {O.}~\bibnamefont {Schmidt}}, \ and\ \bibinfo {author} {\bibfnamefont
  {S.}~\bibnamefont {De~Franceschi}},\ }\href@noop {} {\bibfield  {journal}
  {\bibinfo  {journal} {Appl. Phys. Lett.}\ }\textbf {\bibinfo {volume}
  {103}},\ \bibinfo {pages} {263113} (\bibinfo {year} {2013})}\BibitemShut
  {NoStop}%
\bibitem [{\citenamefont {Voisin}\ \emph {et~al.}(2016)\citenamefont {Voisin},
  \citenamefont {Maurand}, \citenamefont {Barraud}, \citenamefont {Vinet},
  \citenamefont {Jehl}, \citenamefont {Sanquer}, \citenamefont {Renard},\ and\
  \citenamefont {De~Franceschi}}]{VoisinNL}%
  \BibitemOpen
  \bibfield  {author} {\bibinfo {author} {\bibfnamefont {B.}~\bibnamefont
  {Voisin}}, \bibinfo {author} {\bibfnamefont {R.}~\bibnamefont {Maurand}},
  \bibinfo {author} {\bibfnamefont {S.}~\bibnamefont {Barraud}}, \bibinfo
  {author} {\bibfnamefont {M.}~\bibnamefont {Vinet}}, \bibinfo {author}
  {\bibfnamefont {X.}~\bibnamefont {Jehl}}, \bibinfo {author} {\bibfnamefont
  {M.}~\bibnamefont {Sanquer}}, \bibinfo {author} {\bibfnamefont
  {J.}~\bibnamefont {Renard}}, \ and\ \bibinfo {author} {\bibfnamefont
  {S.}~\bibnamefont {De~Franceschi}},\ }\href@noop {} {\bibfield  {journal}
  {\bibinfo  {journal} {Nano Lett.}\ }\textbf {\bibinfo {volume} {16}},\
  \bibinfo {pages} {88} (\bibinfo {year} {2016})}\BibitemShut {NoStop}%
\bibitem [{\citenamefont {Nowack}\ \emph {et~al.}(2007)\citenamefont {Nowack},
  \citenamefont {Koppens}, \citenamefont {Nazarov},\ and\ \citenamefont
  {Vandersypen}}]{Nowack07}%
  \BibitemOpen
  \bibfield  {author} {\bibinfo {author} {\bibfnamefont {K.}~\bibnamefont
  {Nowack}}, \bibinfo {author} {\bibfnamefont {F.}~\bibnamefont {Koppens}},
  \bibinfo {author} {\bibfnamefont {Y.~V.}\ \bibnamefont {Nazarov}}, \ and\
  \bibinfo {author} {\bibfnamefont {L.}~\bibnamefont {Vandersypen}},\
  }\href@noop {} {\bibfield  {journal} {\bibinfo  {journal} {Science}\ }\textbf
  {\bibinfo {volume} {318}},\ \bibinfo {pages} {1430} (\bibinfo {year}
  {2007})}\BibitemShut {NoStop}%
\bibitem [{\citenamefont {Nadj-Perge}\ \emph {et~al.}(2010)\citenamefont
  {Nadj-Perge}, \citenamefont {Frolov}, \citenamefont {Bakkers},\ and\
  \citenamefont {Kouwenhoven}}]{LPKSOqubit}%
  \BibitemOpen
  \bibfield  {author} {\bibinfo {author} {\bibfnamefont {S.}~\bibnamefont
  {Nadj-Perge}}, \bibinfo {author} {\bibfnamefont {S.}~\bibnamefont {Frolov}},
  \bibinfo {author} {\bibfnamefont {E.}~\bibnamefont {Bakkers}}, \ and\
  \bibinfo {author} {\bibfnamefont {L.~P.}\ \bibnamefont {Kouwenhoven}},\
  }\href@noop {} {\bibfield  {journal} {\bibinfo  {journal} {Nature}\ }\textbf
  {\bibinfo {volume} {468}},\ \bibinfo {pages} {1084} (\bibinfo {year}
  {2010})}\BibitemShut {NoStop}%
\bibitem [{\citenamefont {Petersson}\ \emph {et~al.}(2012)\citenamefont
  {Petersson}, \citenamefont {McFaul}, \citenamefont {Schroer}, \citenamefont
  {Jung}, \citenamefont {Taylor}, \citenamefont {Houck},\ and\ \citenamefont
  {Petta}}]{PeterssonQED}%
  \BibitemOpen
  \bibfield  {author} {\bibinfo {author} {\bibfnamefont {K.}~\bibnamefont
  {Petersson}}, \bibinfo {author} {\bibfnamefont {L.}~\bibnamefont {McFaul}},
  \bibinfo {author} {\bibfnamefont {M.}~\bibnamefont {Schroer}}, \bibinfo
  {author} {\bibfnamefont {M.}~\bibnamefont {Jung}}, \bibinfo {author}
  {\bibfnamefont {J.~M.}\ \bibnamefont {Taylor}}, \bibinfo {author}
  {\bibfnamefont {A.~A.}\ \bibnamefont {Houck}}, \ and\ \bibinfo {author}
  {\bibfnamefont {J.}~\bibnamefont {Petta}},\ }\href@noop {} {\bibfield
  {journal} {\bibinfo  {journal} {Nature}\ }\textbf {\bibinfo {volume} {490}},\
  \bibinfo {pages} {380} (\bibinfo {year} {2012})}\BibitemShut {NoStop}%
\bibitem [{\citenamefont {Van~den Berg}\ \emph {et~al.}(2013)\citenamefont
  {Van~den Berg}, \citenamefont {Nadj-Perge}, \citenamefont {Pribiag},
  \citenamefont {Plissard}, \citenamefont {Bakkers}, \citenamefont {Frolov},\
  and\ \citenamefont {Kouwenhoven}}]{SOfastLPK}%
  \BibitemOpen
  \bibfield  {author} {\bibinfo {author} {\bibfnamefont {J.}~\bibnamefont
  {Van~den Berg}}, \bibinfo {author} {\bibfnamefont {S.}~\bibnamefont
  {Nadj-Perge}}, \bibinfo {author} {\bibfnamefont {V.}~\bibnamefont {Pribiag}},
  \bibinfo {author} {\bibfnamefont {S.}~\bibnamefont {Plissard}}, \bibinfo
  {author} {\bibfnamefont {E.}~\bibnamefont {Bakkers}}, \bibinfo {author}
  {\bibfnamefont {S.}~\bibnamefont {Frolov}}, \ and\ \bibinfo {author}
  {\bibfnamefont {L.}~\bibnamefont {Kouwenhoven}},\ }\href@noop {} {\bibfield
  {journal} {\bibinfo  {journal} {Phys. Rev. Lett.}\ }\textbf {\bibinfo
  {volume} {110}},\ \bibinfo {pages} {066806} (\bibinfo {year}
  {2013})}\BibitemShut {NoStop}%
\bibitem [{\citenamefont {Levitov}\ and\ \citenamefont
  {Rashba}(2003)}]{Levitov}%
  \BibitemOpen
  \bibfield  {author} {\bibinfo {author} {\bibfnamefont {L.~S.}\ \bibnamefont
  {Levitov}}\ and\ \bibinfo {author} {\bibfnamefont {E.~I.}\ \bibnamefont
  {Rashba}},\ }\href@noop {} {\bibfield  {journal} {\bibinfo  {journal} {Phys.
  Rev. B}\ }\textbf {\bibinfo {volume} {67}},\ \bibinfo {pages} {115324}
  (\bibinfo {year} {2003})}\BibitemShut {NoStop}%
\bibitem [{\citenamefont {Debald}\ and\ \citenamefont {Emary}(2005)}]{Debald}%
  \BibitemOpen
  \bibfield  {author} {\bibinfo {author} {\bibfnamefont {S.}~\bibnamefont
  {Debald}}\ and\ \bibinfo {author} {\bibfnamefont {C.}~\bibnamefont {Emary}},\
  }\href@noop {} {\bibfield  {journal} {\bibinfo  {journal} {Phys. Rev. Lett.}\
  }\textbf {\bibinfo {volume} {94}},\ \bibinfo {pages} {226803} (\bibinfo
  {year} {2005})}\BibitemShut {NoStop}%
\bibitem [{\citenamefont {Flindt}\ \emph {et~al.}(2006)\citenamefont {Flindt},
  \citenamefont {S{\o}rensen},\ and\ \citenamefont {Flensberg}}]{FlensbergSO}%
  \BibitemOpen
  \bibfield  {author} {\bibinfo {author} {\bibfnamefont {C.}~\bibnamefont
  {Flindt}}, \bibinfo {author} {\bibfnamefont {A.~S.}\ \bibnamefont
  {S{\o}rensen}}, \ and\ \bibinfo {author} {\bibfnamefont {K.}~\bibnamefont
  {Flensberg}},\ }\href@noop {} {\bibfield  {journal} {\bibinfo  {journal}
  {Phys. Rev. Lett.}\ }\textbf {\bibinfo {volume} {97}},\ \bibinfo {pages}
  {240501} (\bibinfo {year} {2006})}\BibitemShut {NoStop}%
\bibitem [{\citenamefont {Golovach}\ \emph {et~al.}(2006)\citenamefont
  {Golovach}, \citenamefont {Borhani},\ and\ \citenamefont {Loss}}]{Golovach}%
  \BibitemOpen
  \bibfield  {author} {\bibinfo {author} {\bibfnamefont {V.~N.}\ \bibnamefont
  {Golovach}}, \bibinfo {author} {\bibfnamefont {M.}~\bibnamefont {Borhani}}, \
  and\ \bibinfo {author} {\bibfnamefont {D.}~\bibnamefont {Loss}},\ }\href@noop
  {} {\bibfield  {journal} {\bibinfo  {journal} {Phys. Rev. B}\ }\textbf
  {\bibinfo {volume} {74}},\ \bibinfo {pages} {165319} (\bibinfo {year}
  {2006})}\BibitemShut {NoStop}%
\bibitem [{\citenamefont {Prati}\ \emph {et~al.}(2012)\citenamefont {Prati},
  \citenamefont {De~Michielis}, \citenamefont {Belli}, \citenamefont {Cocco},
  \citenamefont {Fanciulli}, \citenamefont {Kotekar-Patil}, \citenamefont
  {Ruoff}, \citenamefont {Kern}, \citenamefont {Wharam}, \citenamefont
  {Verduijn}, \citenamefont {Tettamanzi}, \citenamefont {Rogge}, \citenamefont
  {Roche}, \citenamefont {Wacquez}, \citenamefont {Jehl}, \citenamefont
  {Vinet},\ and\ \citenamefont {Sanquer}}]{afsid}%
  \BibitemOpen
  \bibfield  {author} {\bibinfo {author} {\bibfnamefont {E.}~\bibnamefont
  {Prati}}, \bibinfo {author} {\bibfnamefont {M.}~\bibnamefont {De~Michielis}},
  \bibinfo {author} {\bibfnamefont {M.}~\bibnamefont {Belli}}, \bibinfo
  {author} {\bibfnamefont {S.}~\bibnamefont {Cocco}}, \bibinfo {author}
  {\bibfnamefont {M.}~\bibnamefont {Fanciulli}}, \bibinfo {author}
  {\bibfnamefont {D.}~\bibnamefont {Kotekar-Patil}}, \bibinfo {author}
  {\bibfnamefont {M.}~\bibnamefont {Ruoff}}, \bibinfo {author} {\bibfnamefont
  {D.}~\bibnamefont {Kern}}, \bibinfo {author} {\bibfnamefont {D.}~\bibnamefont
  {Wharam}}, \bibinfo {author} {\bibfnamefont {J.}~\bibnamefont {Verduijn}},
  \bibinfo {author} {\bibfnamefont {G.}~\bibnamefont {Tettamanzi}}, \bibinfo
  {author} {\bibfnamefont {S.}~\bibnamefont {Rogge}}, \bibinfo {author}
  {\bibfnamefont {B.}~\bibnamefont {Roche}}, \bibinfo {author} {\bibfnamefont
  {R.}~\bibnamefont {Wacquez}}, \bibinfo {author} {\bibfnamefont
  {X.}~\bibnamefont {Jehl}}, \bibinfo {author} {\bibfnamefont {M.}~\bibnamefont
  {Vinet}}, \ and\ \bibinfo {author} {\bibfnamefont {M.}~\bibnamefont
  {Sanquer}},\ }\href@noop {} {\bibfield  {journal} {\bibinfo  {journal}
  {Nanotechnology}\ }\textbf {\bibinfo {volume} {23}},\ \bibinfo {pages}
  {215204} (\bibinfo {year} {2012})}\BibitemShut {NoStop}%
\bibitem [{\citenamefont {Ono}\ \emph {et~al.}(2002)\citenamefont {Ono},
  \citenamefont {Austing}, \citenamefont {Tokura},\ and\ \citenamefont
  {Tarucha}}]{TaruchaPSB}%
  \BibitemOpen
  \bibfield  {author} {\bibinfo {author} {\bibfnamefont {K.}~\bibnamefont
  {Ono}}, \bibinfo {author} {\bibfnamefont {D.}~\bibnamefont {Austing}},
  \bibinfo {author} {\bibfnamefont {Y.}~\bibnamefont {Tokura}}, \ and\ \bibinfo
  {author} {\bibfnamefont {S.}~\bibnamefont {Tarucha}},\ }\href@noop {}
  {\bibfield  {journal} {\bibinfo  {journal} {Science}\ }\textbf {\bibinfo
  {volume} {297}},\ \bibinfo {pages} {1313} (\bibinfo {year}
  {2002})}\BibitemShut {NoStop}%
\bibitem [{\citenamefont {Kotekar-Patil}\ \emph {et~al.}(2017)\citenamefont
  {Kotekar-Patil}, \citenamefont {Corna}, \citenamefont {Maurand},
  \citenamefont {Crippa}, \citenamefont {Orlov}, \citenamefont {Barraud},
  \citenamefont {Hutin}, \citenamefont {Vinet}, \citenamefont {Jehl},
  \citenamefont {De~Franceschi},\ and\ \citenamefont {Sanquer}}]{DharamPSB}%
  \BibitemOpen
  \bibfield  {author} {\bibinfo {author} {\bibfnamefont {D.}~\bibnamefont
  {Kotekar-Patil}}, \bibinfo {author} {\bibfnamefont {A.}~\bibnamefont
  {Corna}}, \bibinfo {author} {\bibfnamefont {R.}~\bibnamefont {Maurand}},
  \bibinfo {author} {\bibfnamefont {A.}~\bibnamefont {Crippa}}, \bibinfo
  {author} {\bibfnamefont {A.}~\bibnamefont {Orlov}}, \bibinfo {author}
  {\bibfnamefont {S.}~\bibnamefont {Barraud}}, \bibinfo {author} {\bibfnamefont
  {L.}~\bibnamefont {Hutin}}, \bibinfo {author} {\bibfnamefont
  {M.}~\bibnamefont {Vinet}}, \bibinfo {author} {\bibfnamefont
  {X.}~\bibnamefont {Jehl}}, \bibinfo {author} {\bibfnamefont {S.}~\bibnamefont
  {De~Franceschi}}, \ and\ \bibinfo {author} {\bibfnamefont {M.}~\bibnamefont
  {Sanquer}},\ }\href@noop {} {\bibfield  {journal} {\bibinfo  {journal} {Phys.
  S. Sol. (b)}\ }\textbf {\bibinfo {volume} {254}} (\bibinfo {year}
  {2017})}\BibitemShut {NoStop}%
\bibitem [{\citenamefont {Bohuslavskyi}\ \emph {et~al.}(2016)\citenamefont
  {Bohuslavskyi}, \citenamefont {Kotekar-Patil}, \citenamefont {Maurand},
  \citenamefont {Corna}, \citenamefont {Barraud}, \citenamefont {Bourdet},
  \citenamefont {Hutin}, \citenamefont {Niquet}, \citenamefont {Jehl},
  \citenamefont {De~Franceschi},\ and\ \citenamefont {Sanquer}}]{HeorhiiPSB}%
  \BibitemOpen
  \bibfield  {author} {\bibinfo {author} {\bibfnamefont {H.}~\bibnamefont
  {Bohuslavskyi}}, \bibinfo {author} {\bibfnamefont {D.}~\bibnamefont
  {Kotekar-Patil}}, \bibinfo {author} {\bibfnamefont {R.}~\bibnamefont
  {Maurand}}, \bibinfo {author} {\bibfnamefont {A.}~\bibnamefont {Corna}},
  \bibinfo {author} {\bibfnamefont {S.}~\bibnamefont {Barraud}}, \bibinfo
  {author} {\bibfnamefont {L.}~\bibnamefont {Bourdet}}, \bibinfo {author}
  {\bibfnamefont {L.}~\bibnamefont {Hutin}}, \bibinfo {author} {\bibfnamefont
  {Y.-M.}\ \bibnamefont {Niquet}}, \bibinfo {author} {\bibfnamefont
  {X.}~\bibnamefont {Jehl}}, \bibinfo {author} {\bibfnamefont {S.}~\bibnamefont
  {De~Franceschi}}, \ and\ \bibinfo {author} {\bibfnamefont {M.}~\bibnamefont
  {Sanquer}},\ }\href@noop {} {\bibfield  {journal} {\bibinfo  {journal} {Appl.
  Phys. Lett.}\ }\textbf {\bibinfo {volume} {109}},\ \bibinfo {pages} {193101}
  (\bibinfo {year} {2016})}\BibitemShut {NoStop}%
\bibitem [{\citenamefont {Li}\ \emph {et~al.}(2015)\citenamefont {Li},
  \citenamefont {Hudson}, \citenamefont {Dzurak},\ and\ \citenamefont
  {Hamilton}}]{HamiltonNL15}%
  \BibitemOpen
  \bibfield  {author} {\bibinfo {author} {\bibfnamefont {R.}~\bibnamefont
  {Li}}, \bibinfo {author} {\bibfnamefont {F.~E.}\ \bibnamefont {Hudson}},
  \bibinfo {author} {\bibfnamefont {A.~S.}\ \bibnamefont {Dzurak}}, \ and\
  \bibinfo {author} {\bibfnamefont {A.~R.}\ \bibnamefont {Hamilton}},\
  }\href@noop {} {\bibfield  {journal} {\bibinfo  {journal} {Nano Lett.}\
  }\textbf {\bibinfo {volume} {15}},\ \bibinfo {pages} {7314} (\bibinfo {year}
  {2015})}\BibitemShut {NoStop}%
\bibitem [{\citenamefont {Kawakami}\ \emph {et~al.}(2014)\citenamefont
  {Kawakami}, \citenamefont {Scarlino}, \citenamefont {Ward}, \citenamefont
  {Braakman}, \citenamefont {Savage}, \citenamefont {Lagally}, \citenamefont
  {Friesen}, \citenamefont {Coppersmith}, \citenamefont {Eriksson},\ and\
  \citenamefont {Vandersypen}}]{Vandersypen_electronqubit}%
  \BibitemOpen
  \bibfield  {author} {\bibinfo {author} {\bibfnamefont {E.}~\bibnamefont
  {Kawakami}}, \bibinfo {author} {\bibfnamefont {P.}~\bibnamefont {Scarlino}},
  \bibinfo {author} {\bibfnamefont {D.~R.}\ \bibnamefont {Ward}}, \bibinfo
  {author} {\bibfnamefont {F.~R.}\ \bibnamefont {Braakman}}, \bibinfo {author}
  {\bibfnamefont {D.~E.}\ \bibnamefont {Savage}}, \bibinfo {author}
  {\bibfnamefont {M.~G.}\ \bibnamefont {Lagally}}, \bibinfo {author}
  {\bibfnamefont {M.}~\bibnamefont {Friesen}}, \bibinfo {author} {\bibfnamefont
  {S.~N.}\ \bibnamefont {Coppersmith}}, \bibinfo {author} {\bibfnamefont
  {M.~A.}\ \bibnamefont {Eriksson}}, \ and\ \bibinfo {author} {\bibfnamefont
  {L.~M.~K.}\ \bibnamefont {Vandersypen}},\ }\href@noop {} {\bibfield
  {journal} {\bibinfo  {journal} {Nat. Nanotech.}\ }\textbf {\bibinfo {volume}
  {9}},\ \bibinfo {pages} {666} (\bibinfo {year} {2014})}\BibitemShut {NoStop}%
\bibitem [{\citenamefont {Winkler}\ \emph {et~al.}(2003)\citenamefont
  {Winkler}, \citenamefont {Papadakis}, \citenamefont {De~Poortere},\ and\
  \citenamefont {Shayegan}}]{Winkler03}%
  \BibitemOpen
  \bibfield  {author} {\bibinfo {author} {\bibfnamefont {R.}~\bibnamefont
  {Winkler}}, \bibinfo {author} {\bibfnamefont {S.}~\bibnamefont {Papadakis}},
  \bibinfo {author} {\bibfnamefont {E.}~\bibnamefont {De~Poortere}}, \ and\
  \bibinfo {author} {\bibfnamefont {M.}~\bibnamefont {Shayegan}},\ }\href@noop
  {} {\emph {\bibinfo {title} {Spin-Orbit Coupling in Two-Dimensional Electron
  and Hole Systems}}},\ Vol.~\bibinfo {volume} {41}\ (\bibinfo  {publisher}
  {Springer},\ \bibinfo {year} {2003})\BibitemShut {NoStop}%
\bibitem [{Sup()}]{SupplMat}%
  \BibitemOpen
  \href@noop {} {}\bibinfo {note} {See Supplemental Material for details on
  $g$-matrix formalism, reformulation of the model in Ref.\,\cite{Golovach} in
  the $g$-matrix picture and evaluation of $\hat{g}_{\rm TMR}^\prime$ and
  $\hat{g}_{\rm IZR}^\prime$ matrices from experimental data.}\BibitemShut
  {Stop}%
\bibitem [{\citenamefont {Chibotaru}\ \emph {et~al.}(2008)\citenamefont
  {Chibotaru}, \citenamefont {Ceulemans},\ and\ \citenamefont
  {Bolvin}}]{ChibotaruPRL08}%
  \BibitemOpen
  \bibfield  {author} {\bibinfo {author} {\bibfnamefont {L.}~\bibnamefont
  {Chibotaru}}, \bibinfo {author} {\bibfnamefont {A.}~\bibnamefont
  {Ceulemans}}, \ and\ \bibinfo {author} {\bibfnamefont {H.}~\bibnamefont
  {Bolvin}},\ }\href@noop {} {\bibfield  {journal} {\bibinfo  {journal} {Phys.
  Rev. Lett.}\ }\textbf {\bibinfo {volume} {101}},\ \bibinfo {pages} {033003}
  (\bibinfo {year} {2008})}\BibitemShut {NoStop}%
\bibitem [{\citenamefont {Weil}\ and\ \citenamefont
  {Bolton}(2007)}]{Weil-Bolton}%
  \BibitemOpen
  \bibfield  {author} {\bibinfo {author} {\bibfnamefont {J.~A.}\ \bibnamefont
  {Weil}}\ and\ \bibinfo {author} {\bibfnamefont {J.~R.}\ \bibnamefont
  {Bolton}},\ }\href@noop {} {\emph {\bibinfo {title} {Electron paramagnetic
  resonance: elementary theory and practical applications}}}\ (\bibinfo
  {publisher} {John Wiley \& Sons},\ \bibinfo {year} {2007})\BibitemShut
  {NoStop}%
\bibitem [{not()}]{noteU}%
  \BibitemOpen
  \href@noop {} {}\bibinfo {note} {Let $R$ be an unitary transform in the
  $\{\ket{\Uparrow},\ket{\Downarrow}\}$ subspace. $R$ can be cast in the form
  \begin{equation} R=\begin{pmatrix}{} \alpha e^{i\theta} & -\beta^* \\ \beta
  e^{i\theta} & \alpha^* \end{pmatrix} \end{equation} with
  $|\alpha|^2+|\beta|^2=1$. In the basis set
  $\{\ket{\Uparrow}^\prime,\ket{\Downarrow}^\prime\}=R\{\ket{\Uparrow},\ket{\Downarrow}\}$,
  the Hamiltonian reads
  $H^\prime=\mu_B{^t}\vec{\sigma}^\prime\cdot\hat{g}\cdot\vec{B}/2$, with
  $\sigma^\prime_i=R^\dag\sigma_i R$. Yet
  $\vec{\sigma}^\prime=\hat{U}\cdot\vec{\sigma}$, with: \begin{equation}
  \hat{U}=\begin{pmatrix}{} {\rm Re}[(\alpha^2-\beta^2)e^{i\theta}] & {\rm
  Im}[(\alpha^2-\beta^2)e^{i\theta}] & 2{\rm Re}(\alpha^*\beta) \\ -{\rm
  Im}[(\alpha^2+\beta^2)e^{i\theta}] & {\rm Re}[(\alpha^2+\beta^2)e^{i\theta}]
  & 2{\rm Im}(\alpha^*\beta) \\ -2{\rm Re}(\alpha\beta e^{i\theta}) & -2{\rm
  Im}(\alpha\beta e^{i\theta}) & |\alpha|^2-|\beta|^2 \end{pmatrix}\,.
  \label{eqU} \end{equation} Hence
  $H^\prime=\mu_B{^t}\vec{\sigma}\cdot\hat{g}^\prime\cdot\vec{B}/2$, with
  $\hat{g}^\prime={^t}\hat{U}\cdot\hat{g}$. The $\hat{U}$ matrix is unitary
  with determinant $+1$. Therefore, any rotation of the
  $\{\ket{\Uparrow},\ket{\Downarrow}\}$ basis set results in a corresponding
  rotation of the $g$-matrix. Conversely, any unitary $3\times3$ matrix
  $\hat{U}$ with determinant $+1$ can be mapped onto Eq. (\ref{eqU}), and
  associated with a unitary transform $R$ in the
  $\{\ket{\Uparrow},\ket{\Downarrow}\}$ subspace.}\BibitemShut {Stop}%
\bibitem [{\citenamefont {Golovach}\ \emph {et~al.}(2010)\citenamefont
  {Golovach}, \citenamefont {Borhani},\ and\ \citenamefont
  {Loss}}]{Golovach10}%
  \BibitemOpen
  \bibfield  {author} {\bibinfo {author} {\bibfnamefont {V.~N.}\ \bibnamefont
  {Golovach}}, \bibinfo {author} {\bibfnamefont {M.}~\bibnamefont {Borhani}}, \
  and\ \bibinfo {author} {\bibfnamefont {D.}~\bibnamefont {Loss}},\ }\href
  {\doibase 10.1103/PhysRevA.81.022315} {\bibfield  {journal} {\bibinfo
  {journal} {Phys. Rev. A}\ }\textbf {\bibinfo {volume} {81}},\ \bibinfo
  {pages} {022315} (\bibinfo {year} {2010})}\BibitemShut {NoStop}%
\bibitem [{\citenamefont {Corna}\ \emph {et~al.}(2018)\citenamefont {Corna},
  \citenamefont {Bourdet}, \citenamefont {Maurand}, \citenamefont {Crippa},
  \citenamefont {Kotekar-Patil}, \citenamefont {Bohuslavskyi}, \citenamefont
  {Lavieville}, \citenamefont {Hutin}, \citenamefont {Barraud}, \citenamefont
  {Jehl}, \citenamefont {Vinet}, \citenamefont {De~Franceschi}, \citenamefont
  {Niquet},\ and\ \citenamefont {Sanquer}}]{Corna17}%
  \BibitemOpen
  \bibfield  {author} {\bibinfo {author} {\bibfnamefont {A.}~\bibnamefont
  {Corna}}, \bibinfo {author} {\bibfnamefont {L.}~\bibnamefont {Bourdet}},
  \bibinfo {author} {\bibfnamefont {R.}~\bibnamefont {Maurand}}, \bibinfo
  {author} {\bibfnamefont {A.}~\bibnamefont {Crippa}}, \bibinfo {author}
  {\bibfnamefont {D.}~\bibnamefont {Kotekar-Patil}}, \bibinfo {author}
  {\bibfnamefont {H.}~\bibnamefont {Bohuslavskyi}}, \bibinfo {author}
  {\bibfnamefont {R.}~\bibnamefont {Lavieville}}, \bibinfo {author}
  {\bibfnamefont {L.}~\bibnamefont {Hutin}}, \bibinfo {author} {\bibfnamefont
  {S.}~\bibnamefont {Barraud}}, \bibinfo {author} {\bibfnamefont
  {X.}~\bibnamefont {Jehl}}, \bibinfo {author} {\bibfnamefont {M.}~\bibnamefont
  {Vinet}}, \bibinfo {author} {\bibfnamefont {S.}~\bibnamefont
  {De~Franceschi}}, \bibinfo {author} {\bibfnamefont {Y.-M.}\ \bibnamefont
  {Niquet}}, \ and\ \bibinfo {author} {\bibfnamefont {M.}~\bibnamefont
  {Sanquer}},\ }\href@noop {} {\bibfield  {journal} {\bibinfo  {journal} {npj
  Quantum Info.}\ }\textbf {\bibinfo {volume} {4}},\ \bibinfo {pages} {6}
  (\bibinfo {year} {2018})}\BibitemShut {NoStop}%
\bibitem [{\citenamefont {Ivchenko}\ and\ \citenamefont
  {Kiselev}(1998)}]{Ivchenko}%
  \BibitemOpen
  \bibfield  {author} {\bibinfo {author} {\bibfnamefont {E.}~\bibnamefont
  {Ivchenko}}\ and\ \bibinfo {author} {\bibfnamefont {A.}~\bibnamefont
  {Kiselev}},\ }\href@noop {} {\bibfield  {journal} {\bibinfo  {journal} {JETP
  Lett.}\ }\textbf {\bibinfo {volume} {76}},\ \bibinfo {pages} {43} (\bibinfo
  {year} {1998})}\BibitemShut {NoStop}%
\bibitem [{\citenamefont {Tadjine}\ \emph {et~al.}(2017)\citenamefont
  {Tadjine}, \citenamefont {Niquet},\ and\ \citenamefont
  {Delerue}}]{Tadjine17}%
  \BibitemOpen
  \bibfield  {author} {\bibinfo {author} {\bibfnamefont {A.}~\bibnamefont
  {Tadjine}}, \bibinfo {author} {\bibfnamefont {Y.-M.}\ \bibnamefont {Niquet}},
  \ and\ \bibinfo {author} {\bibfnamefont {C.}~\bibnamefont {Delerue}},\
  }\href@noop {} {\bibfield  {journal} {\bibinfo  {journal} {Phys. Rev. B}\
  }\textbf {\bibinfo {volume} {95}},\ \bibinfo {pages} {235437} (\bibinfo
  {year} {2017})}\BibitemShut {NoStop}%
\end{thebibliography}

%

\end{document}